\documentclass[12pt]{article}\usepackage[hyperfootnotes=false]{hyperref}
\usepackage{epsfig}
\usepackage{float}
\usepackage{empheq}

\usepackage{caption}

\usepackage{amsmath}
\usepackage{amssymb}
\usepackage{graphicx}
\newlength{\abstractwidth}
\renewcommand{\thefootnote}{\fnsymbol{footnote}}
\renewcommand{\thanks}[1]{\footnote{#1}} 
\newcommand{\starttext}{
\setcounter{footnote}{0}
\renewcommand{\thefootnote}{\arabic{footnote}}}

\newcommand{\be}{\begin{equation}}
\newcommand{\bea}{\begin{eqnarray}}
\newcommand{\eea}{\end{eqnarray}}
\newcommand{\beq}{\begin{equation}}
\newcommand{\ee}{\end{equation}}

\newcommand*\widefbox[1]{\fbox{\hspace{2em}#1\hspace{2em}}}

\def\eq{&=&}

\def\ra{\rangle}

\def\simleq{\; \raise0.3ex\hbox{$<$\kern-0.75em
\raise-1.1ex\hbox{$\sim$}}\; }
\def\simgeq{\; \raise0.3ex\hbox{$>$\kern-0.75em
\raise-1.1ex\hbox{$\sim$}}\; }

\def\bi{\begin{itemize}}
\def\ei{\end{itemize}}

\def\S{Schwarzschild}
\def\sc{\setcounter{equation}{0}}

\def\dof{degrees of freedom }

\def\CC{{\cal{C}}}

\def\CJ{{\cal{J}}}

\def\CL{{\cal{L}}}

\def\Tr{\rm Tr \it}

\def\bsub{ \begin{subequations}
\begin{empheq}[box=\widefbox]{align}  }
\def\esub{ \end{empheq}
\end{subequations}}

\def\bn{\bigskip \noindent}

  \def\tb{\tilde{\beta}}
  \def\tt{\tilde{T}}
  \def\tk{\tilde{\kappa}}
  
  \def\lf{\left(}
    \def\rg{\right)}
    \def\dr{\dot{\rho}}

\makeatletter
\g@addto@macro\normalsize{%
  \setlength\abovedisplayskip{10pt}
  \setlength\belowdisplayskip{20pt}
  \setlength\abovedisplayshortskip{10pt}
  \setlength\belowdisplayshortskip{20pt}
}
\makeatother

\usepackage{color}

\begin{document}

  
\begin{titlepage}

\rightline{}
\bigskip
\bigskip\bigskip\bigskip\bigskip
\bigskip

\centerline{\Large \bf { Complexity and Newton's Laws }}
\bn

\bigskip
\begin{center}
\bf      Leonard Susskind$^1$  \rm

\bigskip
$^1$ Stanford Institute for Theoretical Physics and Department of Physics, \\
Stanford University,
Stanford, CA 94305-4060, USA \\

\end{center}

\bn

\begin{abstract}
In a recent note \cite{Susskind:2018tei} I argued that the holographic origin of ordinary gravitational attraction is the quantum mechanical tendency for operators to grow under time evolution. In a follow-up \cite{Brown:2018kvn} the claim was tested in the context of the SYK theory and its bulk dual---the theory of near-extremal black holes. In this paper I give  an  improved version of the size-momentum correspondence of \cite{Brown:2018kvn}, and show that Newton’s laws of motion are a consequence. 
Operator size is closely related to complexity. Therefore one may say that gravitational attraction is a manifestation of the tendency for complexity to increase. 

 The improved version of the size-momentum correspondence can be justified by the arguments of Lin, Maldacena, and Zhao  \cite{LMZ} constructing symmetry generators for the approximate symmetries of the SYK model.


\end{abstract}

\end{titlepage}

\starttext \baselineskip=17.63pt \setcounter{footnote}{0}
\tableofcontents

\Large
\sc

\sc
\section{Preliminary Remarks}
 
What is it that takes place in the holographic representation of a  theory when an object in the bulk is gravitationally attracted to a massive body? 
Consider a holographic theory representing a region of empty space. By operating with a simple boundary  operator $\psi,$ a particle can be introduced into the bulk. As the particle moves away from the boundary the operator $\psi$ evolves with time,
\be 
\psi(t) = e^{-iHt} \psi e^{iHt},
\ee
and becomes increasingly complex. If expanded in simple boundary operators the average number of such operators will increase and one says the size of the operator grows. A closely related fact is that the complexity of $\psi(t)$ grows. We might expect that the complexity is a good holographic indicator of how far from the boundary the particle is located. However there is more to the particle than just its location; we may want to know how its momentum or velocity is encoded in the evolving operator $\psi(t).$   The size or complexity is not enough to determine both its distance from the boundary and its momentum.

Let's say that the particle is moving away from the boundary so that the size is increasing. It seems plausible that velocity is related to the rate of change of size. This is oversimplified but it roughly captures the idea that size, and its rate of change, holographically encode the motion of the particle.

Now suppose there is a heavy mass at the center of the bulk region. The gravitational pull of the heavy mass will accelerate the particle away from the boundary. We may expect that the growth of $\psi$---both its size and complexity---will be accelerated relative to the empty case. Thus it is plausible that the holographic representation of gravitational attraction has something to do with the tendency for operators to grow and become more complex \cite{Susskind:2018tei}.  Gravity accelerates that tendency. 

In \cite{Brown:2018kvn} the SYK model and its bulk dual, which in many ways resembles the theory of near-extremal Reissner-Nordstrom (NERN) black holes, provided a testing ground for this hypothesis.  In this paper I will continue the line of reasoning of \cite{Brown:2018kvn}.  A  connection between the evolution of complexity and Newton's second and third laws of motion, as well as Newton's law of attraction, will be derived:
\bi 
\item Newton's second law is summarized by the familiar equation,
\be 
F=ma
\ee
or its generalization,
\be 
F = \frac{dP}{dt}.
\ee
\item  Newton's third law---the law of action and reaction---says that the force exerted by $A$ on $B$ is equal and opposite to the force exerted by $B$ on $A$.
\item  Newton's law of attraction,
\be 
F=\frac{mMG}{r^2}
\ee
\ei

\bn

My arguments are a heuristic mix of quantum information and gravitation and involve some guesswork, but  a more formal basis has been found by Lin, Maldacena, and Zhao  \cite{LMZ}. In section \ref{Sec: Formal} I'll briefly explain the connection insofar as I understand it. 

\bn

\subsection*{Note on size and complexity}

The concept of temperature-dependent size that I will use in this paper is due to Qi and Streicher  \cite{Qi:2018bje}.
Size and complexity are logically different concepts but for reasons that will become clear, over the time period relevant for this paper the two are essentially indistinguishable.
 In order to minimize notation, and to avoid confusing size with entropy,  I will use the symbol $\CC$ to represent both.
The quantitative equivalence of size and complexity continues for times of order  the scrambling time, but by then the connection between size and the motion
of an infalling particle breaks down as the particle reaches the stretched horizon.

\bn

\subsection*{Numerical Coefficients}
Many of the equations in this paper are correct up to numerical factors  
 relating SYK quantities to NERN quantities. These factors   are in-principle  computable using numerical SYK techniques, and depend on the locality parameter $q$. I will use the symbol $\approx$ to indicate that an equation is correct up to such numerical factors.

\sc
\section{Near-Extremal Black Holes}
The bulk dual of the SYK model is usually taken to be a version of the $(1+1)$-dimensional Jackiw-Teitelboim dilaton-gravity system. But that description (of a system with no local degrees of freedom) does not do justice to the spectrum of excitations of the SYK system. In many ways SYK is similar to the long throat of a near-extremal charged black hole whose geometry is approximately $AdS_2\times S^2.$ Unlike pure JT gravity SYK contains matter that can propagate in the throat as it would in the NERN geometry, and the properties of quantum-complexity are not well described by the simple dilaton-gravity system  \cite{Brown:2018bms}. For these reasons I prefer the language of  NERN black holes although no exact SYK/NERN correspondence is known.

\bn

To keep the paper self-contained, in this section I will  review  near-extremal  black holes,  and then in section \ref{Sec: SYK-NERN}, the dictionary relating SYK and near-extremal black holes will be explained.
I will closely follow the discussion of NERN black holes  in  \cite{Brown:2018kvn}.

The metric of the $(3+1)$-dimensional Reissner-Nordstrom  black hole is, 
 \bea 
 ds^2 &=& -f(r)dt^2 + \frac{dr^2}{f(r)} + r^2 d\Omega^2 \cr \cr
 f(r) &=& \left(1-\frac{r_+}{r}\right) \left(1-\frac{r_-}{r}\right) .
 \eea
 The  inner (-) and outer (+) horizons are located at,
  $$r_{\pm} \equiv G M \pm \sqrt{G^2 M^2 - GQ^2}.$$
  
  Define 
  \be 
  (r_+ - r_-) = \delta r.
  \label{Dr}
  \ee
The  temperature  is given by, 
 \begin{equation}
 T = \frac{1}{\beta} = \frac{1}{4\pi}\left( \frac{r_+-r_-}{r_+^2} \right).
\end{equation}
or
\be 
\boxed{
T = \frac{1}{\beta}= \frac{\delta r}{4\pi r_+^2}
}
\label{T(D)}
\ee

The extremal limit is defined by $Q^2 = GM^2$  at which point the horizon radii are equal, $r_+ = r_-.$
Our  interest will be  in near-extremal Reissner-Nordstrom (NERN) black holes, for which
$$\delta r << r_+.$$

 In the NERN limit the temperature is small ($\beta \gg r_+$) and the near-horizon region develops a   `throat' whose length is much longer than $r_+$. The throat is an almost-homogeneous cylinder-like region in which the gravitational field is uniform over a long distance. 

\subsection{The geometry of the throat}
 
\begin{figure}[H]
\begin{center}
\includegraphics[scale=.5]{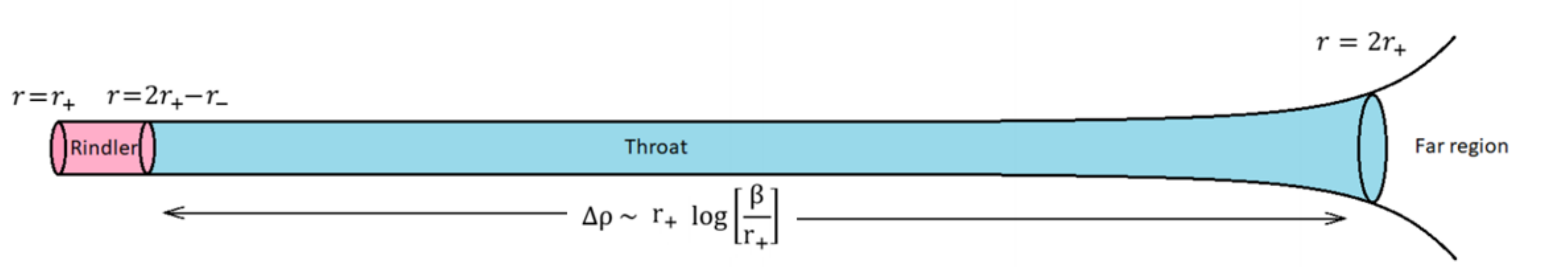}
\caption{The three regions outside a near-extremal charged black hole. Unlike for uncharged black holes, there is now a `throat' separating the Rindler and far regions.}
\label{throatla}
\end{center}
\end{figure}

 The exterior geometry consists of three regions shown  in Fig.~\ref{throatla}.
\begin{itemize}
\item The  \it Rindler  region \rm  is closest to the horizon where the geometry closely resembles the \S \ black hole with the same entropy. It is  defined by,
\be 
r_+ < r  \simleq \   \ 2r_+ - r_-    
\label{are} 
\ee
The Rindler region has proper length $ \sim r_+$ which means that  it's about as long as it is wide.

 The gravitational field (i.e.~the proper acceleration $\alpha = \partial_r \sqrt{f(r)}$ required to remain static at fixed $r$) grows rapidly near the horizon. While the quantity $(1 - \frac{r_+}{r})$ varies significantly in the Rindler region, $(1 - \frac{r_-}{r})$ is essentially constant.

 \bn

\item Proceeding outward, the next region  is the \it throat\rm \   defined by
\be 
2r_+ - r_-  \ \simleq \  r \ \simleq  \ 2r_+
\ee
The throat is long and of almost constant width. The geometry in the throat region is approximately  AdS$_2 \times S^2$, and the gravitational field is almost constant.  The throat ends at $r=2r_+$, which we will soon see is the location of a potential barrier which separates the throat from the far region.
The throat is a feature of charged black holes and is absent from the \S \ black hole. 

For most purposes the geometry in the throat can be approximated by the extremal geometry with $r_+ = r_-.$

The proper length of the throat is,
$$
\Delta \rho  \int_{2r_+-r_-}^{2r_+}\frac{dr}{\sqrt{f}} 
$$
giving

\be 
 \boxed{\Delta \rho  =  r_+  \log\lf {\frac{2\pi \beta}{r_+}}\rg.  }
\ee

We will assume that $ \log\lf {\frac{2\pi \beta}{r_+}}\rg>>1$ which means that the throat is much longer than it is wide.

\bn

\item Next is the \it far region\rm \  where $$(1 - \frac{r_-}{r}) \sim (1 - \frac{r_+}{r}) \sim 1.$$ The far region lies beyond $r=2r_+$. The far region will not be of much interest to us. We will cut it off and replace it by a boundary condition at $r=2 r_+.$
\end{itemize}

\subsection{The black hole boundary}
 
The black hole is effectively sealed off from the far region by a potential barrier.
Low energy quanta in the throat are reflected back as they try to cross from the throat to the far region, or from the far region to the throat.   The barrier height for a NERN black hole is much higher than the temperature and provides a natural boundary of the black hole region. It  may be thought of as the holographic boundary in a quantum description. 
It is also the so-called Schwarzian boundary  that appears  in current literature on SYK theory   \cite{Maldacena:2016hyu}\cite{Maldacena:2016upp}\cite{Maldacena:2017axo}. The boundary will play an important role in this paper.

 The S-wave potential barrier has the form  $$ V(r) = \frac{\partial_r(f^2)}{4r}$$
and for a NERN black hole it is given by,
\be 
\boxed{
V(r)
=
\frac{r_+(r-r_+)^3}{r^6}.
}
\ee
 The width of the barrier in proper distance units is of order $r_+$ and for near extremal RN it is much narrower than the length of the throat. It therefore forms a fairly sharp boundary separating the black hole from the the rest of space.

 At the top  of the barrier  the potential is,
 \be 
 \boxed{
 V_{top} = \left( \frac{1}{8 r_+}  \right)^2 \approx \CJ^2
 \label{potential}
 }
 \ee
 where $\CJ$ is the scale of energy in the SYK theory   (see section \ref{Sec: SYK-NERN}).
 The units of $V$ are energy-squared rather than energy. For a particle to get over the barrier (without tunneling) its energy must be at least $\sqrt{V_{top}}$.
 This is much higher than the thermal scale and for that reason the barrier is very effective at decoupling the black hole, including its thermal atmosphere,  from the far region. Another relevant point is that a particle that starts at rest at the top of the potential has energy of order  $\frac{1}{r_+}\approx \CJ.$ 

The top of the potential barrier serves as an effective boundary of the black hole. It occurs at,
\be 
r = 2 r_+ 
\ee 
We may eliminate reference to the entire region beyond the boundary and replace it by a suitable boundary condition\footnote{In the SYK literature the corresponding boundary condition is placed on the point where the  dilaton achieves a certain value. In the correspondence between the dilaton theory and the NERN black hole the dilaton is simply the area of the local 2-sphere at a given radial location. } on the time-like surface at which $r = 2r_+$. This is accomplished by the introduction of a  boundary term in the gravitational action.

\bn

We   define a radial proper-length coordinate $\rho$ measured from the the black hole boundary\footnote{Frequently a radial proper coordinate is defined as the distance to the horizon. Note that in this paper $\rho$ measures distance to the black hole boundary at $r=2r_+$, not to the horizon.},
\be 
\rho = \int_r^{r_b}\frac{dr'}{\sqrt{f(r'})}
\label{rho}
\ee
In the throat $r$ and $\rho$ are related by,
\be 
\frac{r-r_+}{r_+} = e^{-\rho/r_+}
\label{rho-vs-r}
\ee

At   the boundary $\rho=0,$ and at the beginning of the Rindler region $\rho = r_+ \log{(\beta/r_+)}$. Note that $\rho$ has a large variation over the throat region which makes it a more suitable radial coordinate than $r$ which hardly varies at all.

\bn

The black hole boundary, defined as the place where $r=2r_+,$ is not a rigid immovable object. Fluctuations or dynamical back reaction can change the metric so that the distance from the horizon   to the boundary varies. This can be taken into account  by allowing the boundary to move from its equilibrium position at $\rho = 0.$

\begin{figure}[H]
\begin{center}
\includegraphics[scale=.2]{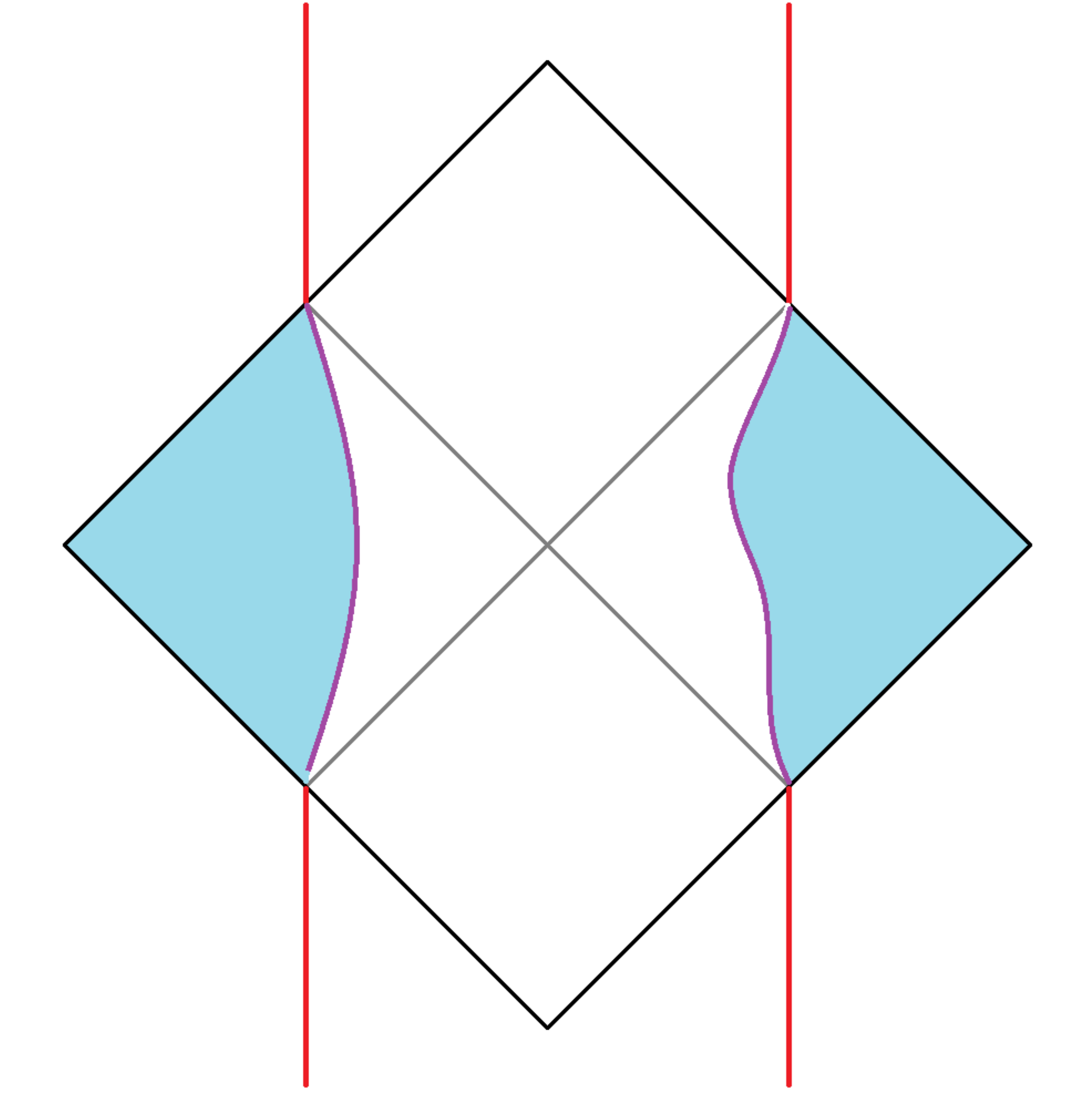}
\caption{Penrose diagram for a NERN black hole. The curved lines represent the trajectory of the black hole boundary at $r=2r_+.$ On the left side the boundary is shown in its equilibrium location while on the right it is moving in reaction to some matter.}
\label{RN-penrose}
\end{center}
\end{figure}

In figure \ref{RN-penrose} a Penrose diagram for a two-sided NERN black hole is shown along with the trajectories of the boundary and the regions beyond the boundary. The left-side boundary is shown in its static equilibrium position but on the right side the dynamical nature of the boundary is illustrated.

 The equation of motion of the boundary is generated by the Hawking-Gibbons-York boundary term (Schwarzian action in SYK literature)  needed to supplement the Einstein-Maxwell action in the presence of a boundary. For small slow perturbations the boundary motion is non-relativistic with a large mass of order 
$S/r_+$  ($S$ is the black hole entropy). The mass of the boundary is of order the mass of the black hole  itself\footnote{The idea of the boundary as a very massive particle was suggested by 
Kitaev, who developed this idea in \cite{Kitaev:2018wpr}. It was further developed in  \cite{Yang:2018gdb}}. Using the SYK-NERN dictionary in section \ref{Sec: SYK-NERN}   we see that the boundary  mass  is,
\be
\boxed{
M_B \approx \CJ N.
}
\ee

\subsection{Particle motion in the throat}
\begin{figure}[H]
\begin{center}
\includegraphics[scale=.4]{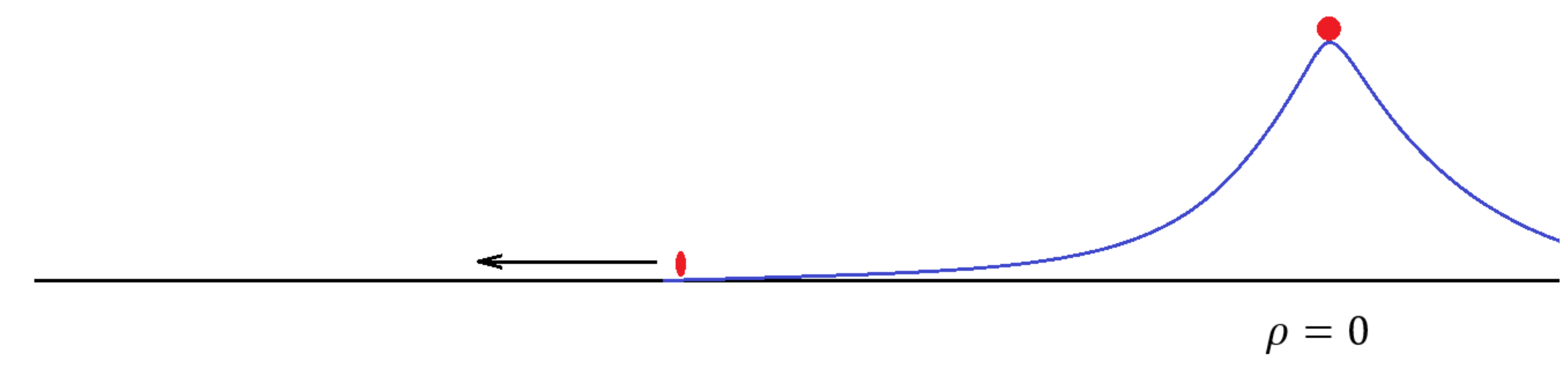}
\caption{A particle is introduced at the top of the potential, and subsequently rolls down the potential,}
\label{fall-from-top}
\end{center}
\end{figure}

Consider a particle dropped at $t=0$ from $\rho =0$, i.e., from the top of the potential as in figure \ref{fall-from-top}. The energy of the particle is $\sim 1/r_+$, which corresponds to an energy $\CJ$ in the SYK theory \cite{Brown:2018kvn}.

Under the influence of a uniform gravitational field it accelerates\footnote{ In the sense that its momentum grows. Being relativistic the velocity is close to $1$. } toward the horizon.  
Appendix \ref{App: Particle Equation of Motion}  works out the equation of motion for the particle, and one finds that the force is constant throughout the throat. The momentum increases linearly with time.
\bn

\begin{figure}[H]
\begin{center}
\includegraphics[scale=.4]{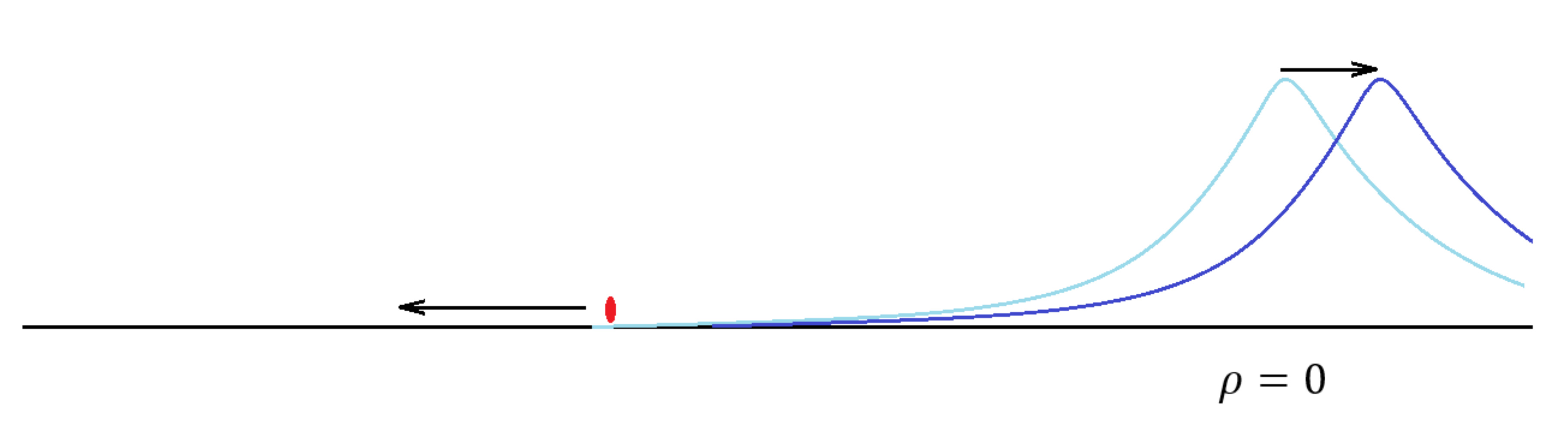}
\caption{The boundary recoils when the particle is accelerated. At all times the particle and the boundary have equal and opposite momentum.}
\label{recoil}
\end{center}
\end{figure}

So far a small but important effect has been ignored.
There is a  back reaction that occurs when the particle falls off the potential. The potential exerts a force on the particle, which in turn exerts an equal and opposite  force on the boundary. The result is that the boundary recoils with a small velocity.
(With some effort this can be seen in the Schwarzian analysis \cite{Maldacena:2017axo}.)  This recoil, illustrated in figure \ref{recoil}, will be important later. 

Once the particle falls off the potential it quickly becomes relativistic. In the throat region its
 trajectory is given by
\bea 
dt \eq \frac{1}{  \sqrt{f}}d\rho \cr \cr
\eq \frac{r_+}{(r-r_+)}d\rho \cr \cr
&=& e^{\rho/r_+}d\rho
\eea
Thus the particle trajectory satisfies,
\be 
t = r_+ (e^{\rho/r_+}  -1)
\label{t of rho}
\ee
or
\be
\boxed{
\rho(t) = r_+\log\lf\frac{t-r_+}{r_+}\rg
}
\label{trajec}
\ee

The total time to fall from $\rho = 0$ to the beginning of the Rindler region is $\beta.$ During that time the distance traveled is 
\be 
\boxed{
\Delta \rho = r_+\log{\lf  \frac{2\pi \beta}{r_+}    \rg}.
}
\ee
\subsection{Schwarzschild $r$ in terms of $\rho$}

Let's consider the relation between the \S \ coordinate $r$ and the proper coordinate $\rho.$
To a very good approximation, in the throat we can assume $r_+ = r_-$ and  that $r$ is constant.  The emblackening factor $$  \lf  \frac{r-r_+}{r} \rg   \lf  \frac{r-r_-}{r} \rg $$ may be replaced by its extremal value
\be 
f(r) \approx \lf  \frac{r-r_+}{r_+} \rg^2  
\ee

Recall that $\rho $ is the proper distance measured from the boundary at $r=2r_+,$ 
\bea 
d\rho \eq\frac{dr}{\sqrt{f(r)}}  \cr \cr
\eq  r_+\frac{dr}{r-r_+}  \cr \cr
\rho \eq  r_+ \log{\lf  \frac{r_+}{r-r_+}  \rg} 
\eea
or,
\be 
\boxed{
\frac{r-r_+}{r_+} = e^{-\rho/r_+}
\label{r-r_+}
}
\ee

\bn

\subsection{Surface gravity and $  \tb  $}
The so-called surface gravity $\kappa$ will play an important role in what follows.  
At the horizon the surface gravity is related to the temperature of the black hole by,
\be 
T = \frac{1}{2\pi} \  \kappa_{horizon}.
\ee 
More generally it is defined at any radial position $r$ by
\be 
\tk(r)= \frac{1}{2} \frac{df}{dr} = \frac{r_+(r-r_-) +r_-(r-r_+)}{2r^3}
\ee
which in the throat is approximated by,
\be 
\tk(r)  = \frac{r-r_+}{r_+^2}
\label{tk}
 \ee
The purpose of the  tilde notation is to indicate a local quantity, i.e., one that may vary throughout the throat. Corresponding variables without the tilde indicate the value of the quantity at the horizon.  We may also define $\tt$ and $\tb$ by,
 \bea
 \tt \eq \frac{1}{2\pi}\tk = \frac{1}{2\pi}\frac{r-r_+}{r_+^2}  \cr  \cr
 \tb \eq \frac{1}{\tt} = 2\pi \frac{r_+^2}{r-r_+}
 \label{tildes}
 \eea
 (Except at the horizon the quantity $\tt$ is not  a real temperature. It is a useful quantity defined by \ref{tk} and \ref{tildes} whose importance will become clear.)
 
 In the throat let's express $\tb$ in terms of $\rho$. Using \ref{trajec}, \ref{r-r_+} and \ref{tildes},
$$
 \tt(\rho) = \frac{1}{2\pi r_+} e^{-\rho/r_+} 
$$
and,
\be 
\boxed{
 \tb(\rho) = 2\pi \ r_+ e^{\rho/r_+} .
 }
 \label{B-of-rho}
 \ee

At $\rho=0,$  \ $\tb$ is given by
\be 
\tb =2\pi \  r_+ \approx  \CJ^{-1}  \ \  \ \ \ \ \ \ \ (\rho = 0)
\label{lim1}
\ee

At the Rindler  end of the throat where $\rho=r_+\log{(\beta/r_+)},$ $\tb$ is given by
 \be 
 \tb = \beta \ \ \ \ \ \ \ \ \ \ \ \ \ \ \ \ \ \ (\rho = \beta)
 \label{lim2}
 \ee

By following the trajectory of the infalling particle \ref{t of rho}, and using \ref{B-of-rho}  we find that $\tb$ grows according to,
\be 
\boxed{
 \tb(t) = 2\pi (t + r_+)
}
\label{B-of-t}
\ee

As the the Rindler region is approached $\tb$ stops increasing and remains at $\beta$ until the horizon is reached.

\sc
\section{SYK/NERN Dictionary}\label{Sec: SYK-NERN}

We can only go so far in  understanding the quantum mechanics of NERN black holes without having a concrete holographic system to analyze. That brings us to the well-studied SYK model. In this section the SYK/NERN dictionary is spelled out. 

\subsection{Qualitative Considerations}
We'll begin with qualitative aspects of the SYK/NERN dictionary and then attempt to determine more precise numerical coefficients in the next subsection.
The two-sided arrows in this subsection  indicate qualitative correspondences..

\bi
\item The overall energy scale of the SYK model is called $\CJ.$ Its inverse $\frac{1}{\CJ}$  is a length scale which corresponds to the \S \ radius $r_+.$  In the SYK model acting with a fermion operator $\psi$ adds an energy $\approx \CJ$. On the NERN side dropping a particle from the top of the barrier adds energy $\approx 1/r_+.$ Thus it makes sense to identify the process of dropping a particle from the black hole boundary, with acting with a single fermion operator. 

\be 
\boxed{
1/r_+ \approx \CJ
\label{1/r+<->J}
}
\ee
%

\item  A single boundary fermion operator in SYK has size $1$ corresponding the the assumption of  \cite{Brown:2018kvn} that the initial size of the operator that creates the particle at the top of the barrier is also $1$.
\be 
\boxed{
\rm size \ of \ 1 \ fermion \leftrightarrow size \ of \ initial \ particle.
}
\label{size=size} 
\ee
\item Up to a numerical factor $\approx 1,$ the zero temperature extremal entropy of  SYK is the number of fermion \dof $N.$
\be 
\boxed{
S_0 \approx N
\label{S0<->N}
}
\ee
%
\item The 4-dimensional Newton constant can be obtained from the entropy formula, $$S_0 = \pi r_+^2 /G $$ Using \ref{1/r+<->J} and \ref{S0<->N} gives,
\be
 \boxed{
G \approx \frac{1}{\CJ^2N}
}
\ee
%
\item The SYK theory does not have sub-AdS locality (locality on scales smaller than $r_+$). It is comparable to a string theory in which the string scale is of order $r_+$ or $1/\CJ.$

\item The black hole mass is $r_+/G.$ This translates to,
\be 
\boxed{
M \approx N \CJ.
}
\label{M<->NJ}
\ee
%
\item Many of the detailed coefficients that appear in the subsequent formulas are dependent on $q,$ the SYK-locality parameter that determines the number of fermion operators in each term in the Hamiltonian. For the most part I will treat $q$ as a constant of order unity and not try to track the $q$-dependent details.

\ei

The literature on the bulk dual of SYK theory \cite{Maldacena:2016upp}\cite{Maldacena:2017axo}\cite{Yang:2018gdb} has its own conventions and notations which are not the standard ones used for  NERN black holes.  Here I'll add to the dictionary the translation between the two.

\bi  
\item 

The dynamical boundary of SYK (described by the Schwarzian action) corresponds to the NERN black hole boundary, i.e., the top of the barrier where the throat meets the far region. The action governing the motion of the boundary is the Gibbons-Hawking-York boundary action.
\be 
\boxed{
 \rm GHY   \leftrightarrow Schwarzian \it 
 }
\label{GHY<->Schw}
\ee

\item
The dilaton field $\phi$ in \cite{Maldacena:2016upp}\cite{Maldacena:2017axo}\cite{Yang:2018gdb} is related to the area of the transverse geometry at a given radial position,
\be 
\phi =4\pi r^2.
\ee

\item The time coordinate  used in the SYK literature is called $u$. It is the  proper time measured at the boundary. We may identify it with the proper time at the top of the potential barrier at $r=2r_+.$

The time coordinate $t$ used in this paper is the asymptotic \S \ time coordinate for the NERN black hole. The relation between $u$ and $t$ is,
\be 
f(r)|_{2r_+}  \ dt^2 = du^2.
\ee
For NERN black holes $f(r)|_{2r_+} = 1/4,$ from which it follows that,
\be 
t=2u.
\label{t=2u}
\ee

\ei

\subsection{Quantitative Considerations}
In some cases the numerical coefficients appearing in the various correspondences have been studied and allow more quantitative correspondences. I'll give some examples here, but I won't keep track of these coefficients in  subsequent sections.

The specific heats of the SYK model and the NERN black hole can both be computed. On the NERN side the calculation is analytic and yeilds,
\be 
c = \frac{dM}{dT} = \frac{4\pi^2}{G}r_+^2 T
\label{c-nern}
\ee
For SYK the  calculation was done in \cite{Maldacena:2016hyu}. The result is,
\be  
c= 4\pi^2 \alpha_s(q) \frac{N}{\CJ}  T
\label{c-syk}
\ee
where $\alpha_s(q)$ is a numerically computed function of the SYK locality parameter $q.$ For $q=4$ $\alpha_s = .007$ and for large $q$ it decreases  $ \sim 1/q^2.$ 

Setting \ref{c-nern} and \ref{c-syk} equal, we find the relation,
\be 
\boxed{
\alpha_s\frac{N}{\CJ} = \frac{r_+^3}{G}.
}
\label{anj}
\ee

Let $\lambda$ and $p$ be dimensionless coefficients defined by,
\be 
G=\frac{\lambda}{\CJ^2 N}
\label{lambda}
\ee
and
\be
r_+ = \frac{p}{\CJ}.
\label{p}
\ee
Plugging \ref{lambda} and \ref{p} into \ref{anj} gives one relation between $p$ and $\lambda,$
\be 
\boxed{
\alpha_s = \frac{p^3}{\lambda}.
}
\label{relat1}
\ee

Another relation can be found by considering the entropy of SYK and the NERN black hole. On the 
NERN side we use the Bekenstein-Hawking formula which  gives,
\be 
S = \frac{\pi r_+^2}{G}.
\label{Sbh}
\ee
On the SYK side reference \cite{Maldacena:2016hyu} Stanford and Maldacena  computed 
the near extremal entropy:
\be 
S= d(q) N.
\label{23}
\ee
where $d(q)$ is another numerically computed function of $q$ which varies from $d(4) =.23$ to $d(\infty) = .35.$

Combing \ref{Sbh} and \ref{23}   with \ref{lambda} and \ref{p} gives another equation for $p$ and $\lambda,$
\be 
\frac{\pi p^2}{\lambda} = d.
\label{relat2}
\ee
The two relations \ref{relat1} and \ref{relat2} yield the following expressions for $\lambda$ and $p,$
\bea 
\lambda \eq  \frac{\pi^3 \alpha_s^2}{d^3}\cr \cr
p\eq \frac{\pi \alpha_s}{d}
\label{lambda&p}
\eea

Thus we find the following  correspondences,
\be 
r_+ =\lf \frac{\pi \alpha_s}{d} \rg \frac{1}{\CJ}
\label{r_+}
\ee

\bn

\be 
G =\lf  \frac{\pi^3 \alpha_s^2}{d^3} \rg \frac{1}{N\CJ^2}.
\label{gee}
\ee

For $q=4$ the numerical values of $\alpha_s$ and $d$  are,
\bea
\alpha_s \eq .007 \cr \cr
d \eq .23
\label{nums}
\eea
giving,
\be 
\boxed{
r_+ = \frac{.10}{\CJ}
}
\label{numr}
\ee
and
\be 
\boxed{
G=\frac{.12}{N\CJ^2}
}
\label{numG}
\ee

Now let's return to the problem of a light particle dropped from the top of the potential \ref{potential} and estimate its energy $\epsilon.$. The height of the barrier is
$$
\sqrt{V_{top} } = 1/8r_+ \approx  \CJ.
$$

We may compare this energy with the energy added to the SYK ground state by applying a single fermion operator $\psi$ (In other words it is the energy associated with a size $1$ perturbation). This energy  is expected to be of order $\CJ$ and to have some smooth $q$ dependence. It is given by,

\be 
\epsilon(q) \CJ= \left\langle    \frac{1}{Z(\beta)}  \Tr H(2\psi e^{-\beta H} \psi - e^{-\beta H})
\right\rangle
\ee
where the average $\left\langle....\right\rangle$ indicates disorder average. (The factor of $2$ in the first term is present because of the SYK convention that $\psi^2 = 1/2.$)

\sc
\section{Growth of Size}

\begin{figure}[H]
\begin{center}
\includegraphics[scale=.4]{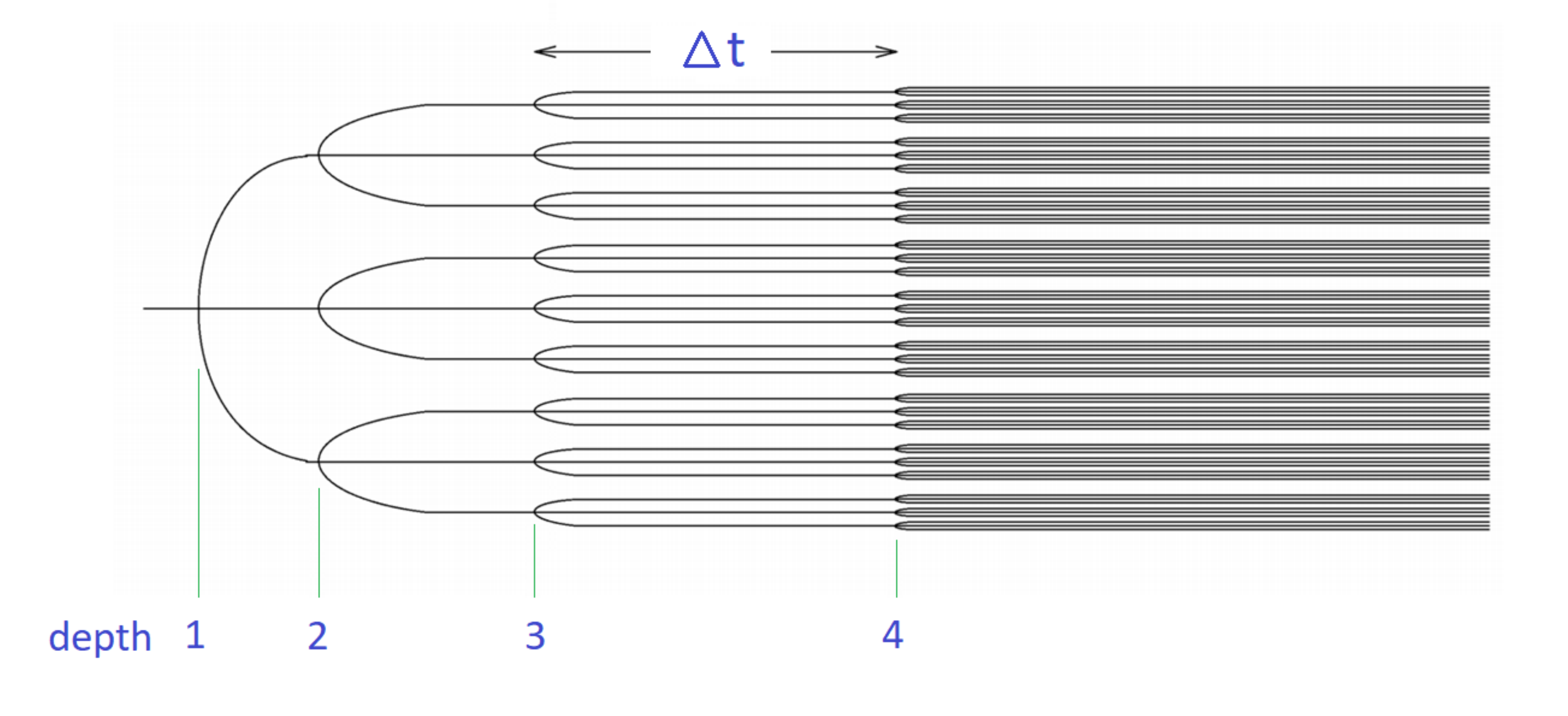}
\caption{Tree-like operator growth. The size at at any circuit-depth is the final number of fermions while the complexity is the number of vertices in the diagram.  In this figure the size is $81$ and the complexity is $40$.  The complexity at the next step would be $40+81 = 121.$
The time scale for a unit change in depth is $\Delta t.$ In general $\Delta t$ may itself be time dependent.}
 \label{oak}
\end{center}
\end{figure}

 Consider applying a single fermion operator at time $t=0.$ The operator evolves in time according to,
\be 
\psi(t) = e^{-iHt}\psi e^{iHt}.
\ee
and  becomes  a superposition of many-fermion operators \cite{Roberts:2018mnp}\cite{Qi:2018bje}. The average number of Fermions at time $t$ is the size.
The evolution is described by Feynman-like diagrams which, up to the scramblinng time, grow exponentially \cite{Roberts:2018mnp}\cite{Qi:2018bje}.  At each stage the average number of fermions increases by  common factor. The process resembles an exponentially expanding tree as shown in figure \ref{oak}.

It is similar to the evolution of a quantum circuit and it is natural to define a circuit depth. In general the circuit depth may not unfold uniformly with time. For example, if for some reason the computer runs at a variable time-dependent rate, the size will grow exponentially with depth but not necessarily with time. The time associated with a unit change in circuit depth is defined to be $\Delta t$ and it may be time-dependent. This type of time-dependence occurs in the evolution of size at low temperature \cite{Qi:2018bje}.

We can express this in terms of a rate of growth $R$,
\bea
R(t) &\equiv& \frac{d\log{\CC(t)}}{dt} \cr \cr
\eq \frac{1}{\Delta t}
\eea

The exponential growth as a function of circuit depth (for time less than the scrambling time)  is the reason that size and complexity are proportional to each other. One may think of the size 
at a given depth as the number of ``leaves" of the tree, and the complexity as the integrated number of vertices up to that point. Because the tree
grows exponentially, the number of leaves and the number of vertices are proportional, and with some normalization (of complexity) the size and complexity can be set equal.

\subsection{Infinite Temperature} 
Roberts, Stanford, and Streicher \cite{Roberts:2018mnp}  have calculated the time dependence of size at infinite temperature and find,
\be 
 \CC(t) \sim e^{2\CJ t}.
\ee
Roberts, Stanford, and Streicher give a more detailed formula,
\be 
\CC(t) = 1 +2 \sinh^2{    \lf  \CJ t  \rg             }
\label{RobStanStreicher}
\ee
Apart from a brief transient the size grows exponentially. Dropping the $1$ which is unimportant,
the rate $R(t)$ is
\bea
R(t) \eq \frac{1}{\CC(t)}\frac{d\CC(t)}{dt} \cr \cr
\eq  \CJ    \frac{\cosh{ \CJ t}}{\sinh{ \CJ t}} 
\eea
which after a short time $\CJ^{-1} $ tends to
\be 
R \to \CJ
\ee

We may restate this in terms of $\Delta t,$
\be 
\Delta t \approx  \CJ^{-1}    \ \ \ \ \ \ \ \ \ (T=\infty).
\ee

\subsection{Low Temperature, $\rm{T<<J}$}
At very low temperatures the pattern
 is quantitatively different.  According to Qi and Streicher the  size for low $T$ is given by,
 \be
\CC(t) = 1+2 \frac{\CJ^2\beta^2}{\pi^2} \sinh^2{\lf  \frac{\pi t}{\beta}  \rg}
\label{QS}
 \ee
Early on the rate is comparable to the  infinite $T$ case,
\be
\boxed{
\frac{1}{\CC} \frac{d\CC}{dt} \approx  \CJ  \ \ \ \ \ \ \ \ \ \ \  \ \  ( \CJ t \sim 1) 
}
\label{limit1}
\ee
but after a time $\beta/2\pi$ (at which the infalling particle has reached the Rindler region) the rate has slowed to 

\be
\boxed{
\frac{1}{\CC} \frac{d\CC}{dt} = \frac{2\pi}{ \beta} \ \ \ \ \ \ \ \ \ \ \  \ \  ( \beta/2\pi < t<t_*) 
}
\label{limit2}
\ee

Our interest will lie  in the throat region during time period  between $t=0$ and $t=2\pi \beta$,  where the rate is time-dependent, varying from $\approx \CJ$ to $2\pi /\beta.$  In fact the rate  is not so much time-dependent as it is position dependent. 
To understand the  the rate in more detail \cite{Qi:2018bje}  we consider a particle falling from the black hole boundary. The particle falls along a trajectory $\rho(t).$ The time dependence of the growth rate is really $\rho$-dependence: the rate depends on $t$ only through the position $\rho$.

Let $\kappa (\rho)$ be the surface gravity at position $\rho$,
\be 
\kappa(\rho) \equiv \frac{1}{2}\partial_r f(r)
\ee
 and let $\tb$ be,
\be 
\tb(\rho) = 2\pi/\kappa(\rho).
\ee

At the horizon the surface gravity is related to the temperature of the black hole,
\be 
T = \frac{1}{2\pi} \  \kappa_{horizon}.
\ee
and $\tb_{horizon}$ to  the inverse temperature,
\be
\tb_{horizon} = \beta 
\ee

\bn

\bn

The obvious guess for the interpolation between \ref{limit1} and \ref{limit2} is,
$$\frac{1}{\CC} \frac{d\CC}{dt} = \frac{2\pi}{  \tb}. $$ This is correct in the Rindler region but in the throat it is off by a factor of $2.$ Consistency between the Qi-Streicher formula and \ref{B-of-t} requires,
\be
\boxed{
\frac{1}{\CC} \frac{d\CC}{dt} \sim \frac{4\pi}{  \tb} 
}
\label{tb-dC/dt=s}
\ee
or in terms of $\Delta t,$
\be 
\Delta t =     \frac{ \tb(\rho)}{4\pi}       .
\label{Dt=tb}
\ee

\sc
\section{Momentum-Size Correspondence}

\subsection{Formulation}
In \cite{Brown:2018kvn} it was proposed that the holographic dual to the momentum of an infalling particle is related to  the size (or complexity) of the operator  that created the particle. By itself this is not dimensionally consistent. One needs a quantity with units of length to multiply the momentum in order to get a dimensionless size. 
For a \S \ black hole there is only one length scale, the \S \ radius, which is proportional  to $\beta/2\pi$.    Thus,
\be 
\CC  \approx  \frac{\beta}{2\pi} P,
\label{s=bp}
\ee
(the factor of proportionality being $q$-dependent).
However in the NERN case this cannot be the right relation. Pick a  point $\rho_0$ a fixed distance from the boundary. If the temperature is sufficiently low the geometry between $\rho=0$ and $\rho=\rho_0$ is extremely  insensitive to $\beta$ and the growth up to that point should also  be insensitive to $\beta$. But equation \ref{s=bp} implies that $\CC(\rho_0)$ blows up as $T\to 0.$

The  formula used in \cite{Brown:2018kvn} was originally suggested  by Ying Zhao.  
It is obtained by replacing  equation \ref{s=bp} 
 by a local version,
\be 
\boxed{
\frac{\tb}{4\pi}           P \approx  \CC.
}
\label{p-SIZE}
\ee

From \ref{p-SIZE} one sees that complexity (or size) is not in one to one relationship with either position ($\rho$) or momentum ($P$) but it is a combination of both variables. For fixed position the complexity is proportional to momentum, but for fixed momentum the complexity increases the deeper the particle is into the throat.
I will not repeat the argument here but just remark that in   \cite{Brown:2018kvn}      it was shown that \ref{p-SIZE}  gives an accurate account of the evolution of size, reproducing a non-trivial result of \cite{Leichenauer:2014nxa}. As we'll now see, it is also agrees with the calculations of  \cite{Qi:2018bje}.

\subsection{Qi-Streicher formula}

 Qi and Streicher  \cite{Qi:2018bje} have made a first-principles calculation of  the growth of a single fermion operator $\psi$ at finite temperature $1/\beta$ in the SYK theory.
As time evolves the complexity of $\psi(t)$ grows until the scrambling time $t_*$. Between $t=0$ and $t=t_*$ Qi and Streicher find\footnote{Qi and Streicher calculate the size but for reasons I have explained size and complexity are interchangeable for our purposes.},
\be
\boxed{
 \CC(t) = 1+2 \frac{\CJ^2\beta^2}{\pi^2} \sinh^2{\lf  \frac{\pi t}{\beta}  \rg}
 }
 \label{Qi-Streicher}
 \ee

\bn
 
 Let us compare \ref{B-of-t}, 
 $$ \tb(t) = 2\pi (t + r_+)$$
 with the SYK calculation of Qi-Streicher. We first note from  \ref{tb-dC/dt=s} that for $t>r_+,$
 \be 
 \frac{2\pi}{ \tb} \sim \frac{d\log{\CC(t)}}{dt}.
 \label{inv-rate}  
 \ee
The first term in the  Qi-Streicher formula \ref{Qi-Streicher} is unimportant.  We may write,
$$ 
\CC(t) = 2 \frac{\CJ^2\beta^2}{\pi^2}\sinh^2{\pi t/\beta}.
$$
and

\be 
\frac{d\log{\CC}}{dt} =\frac{2\pi}{\beta} \tanh^{-1}{\pi t/\beta}.
\ee

Using \ref{inv-rate}   we find,
 \be 
 \boxed{
 \tb \sim \beta \tanh{(\pi t/\beta)}
 }
 \ee
For $r_+ < t < \beta/2\pi$  this gives  $\tb \sim  2\pi  t$  in  agreement with \ref{B-of-t}.

Actually this is accurate for almost the entire passage through the throat. The ratio 
$$  \frac{\beta\tanh{(\pi t/\beta)}}{\pi t} $$
  is close to $1$ as long as $\pi t/\beta <1.$ (Note $ \lf  \frac{\tanh.3}{.3}\rg   = .97$)  In terms of $\rho$ this means until,
  \bea
  \rho &=& r_+ \log{\beta/r_+} - r_+ \log(\pi) \cr \cr
  &=& \Delta \rho  - r_+ \log(\pi) 
  \eea
where $\Delta \rho$ is the length of the throat (see figure \ref{throatla}). 
 In other words there is very good agreement between the Qi-Streicher formula, and the   
rate \ref{tb-dC/dt=s} conjectured in    \cite{Brown:2018kvn},
  over the entire throat,  right up to the start of the Rindler region. The agreement continues to be qualitatively good into the Rindler region. The discrepancy by the time the particle has reached a Planck distance from the horizon is less than a factor of $2$. 
  
  \bn
  
There is a striking similarity between \ref{Qi-Streicher} and the infinite temperature formula \ref{RobStanStreicher} but quantitatively they are quite different. From \ref{RobStanStreicher} we see that at $T=\infty$ the size quickly tends to the exponential form $e^{\CJ t}.$ The quadratic growth only persists for a very short time of order $1/\CJ.$ This shows the lack of a throat region.

By contrast, in the low $T$ limit the quadratic growth last for a time of order $\beta$ which is much greater than $1/\CJ,$ demonstrating the existence of the long throat.

\sc
\section{Newton's Equations for Complexity}\label{Sec: Newt}

\subsection{Complexity and Momentum}
Now we come to the main point, the relation between the evolution of complexity and Newton's equations of motion.
Let us compare  \ref{tb-dC/dt=s},
 $$ 
d\CC \approx   \CC \  \lf \frac{4\pi dt}{\tb(t)} \rg
$$ 
and \ref{p-SIZE},
$$
\frac{\tb}{4\pi}           P \approx  \CC.
$$  
Eliminating $\tb$  we find a  relation\footnote{This relation was   derived by Lin, Maldacena, and Zhao  by different arguments. See  section \ref{Sec: Formal}  and \cite{LMZ}. As in other formulas there is an implicit $q$-dependent proportionality factor. }
\be 
\boxed{
P \approx  \frac{d\CC}{dt}
}
\label{p=dC/dt}
\ee
between a dynamical quantity $P$, and an information-theoretic quantity, complexity: 

\bn
\it
The momentum of an infalling particle created by $\psi$ is proportional to the rate at which the complexity of the precursor $\psi(t)$ grows. 
 \rm

The numerical  constant relating the two sides of \ref{p=dC/dt} is connected with the coefficient $\epsilon$ in the additional energy of applying a fermion operator the SYK low temperature state ground state.

\bn
Equation \ref{p=dC/dt} resembles the ordinary non-relativistic  relation between momentum and velocity. One might be tempted to think that $\frac{d\CC}{dt}$ is proportional to the spatial velocity
of the infalling particle, but the simple proportionality of momentum and velocity is only valid for non-relativistic motion.  The infalling particle however  quickly becomes relativistic.

\bn

Nevertheless let's proceed to time-differentiate [\ref{p=dC/dt}],
\be 
\frac{dP}{dt}\approx  \frac{d^2\CC}{dt^2}.
\label{dp/dt=F}
\ee
We next use the fact that the rate of change  of  momentum is the applied  force,
\be 
\boxed{
F \approx   \frac{d^2\CC}{dt^2}.
}
\label{FC}
\ee

In appendix \ref{App: Particle Equation of Motion} the force $F$ on an infalling particle in the gravitational field of a NERN black hole is calculated using the standard Lagrangian formulation of particle mechanics. It is explicitly shown to agree with
 $ \frac{d^2\CC}{dt^2}$ 
 as calculated from the Qi-Streicher formula---the  formula being a pure SYK relation whose derivation does not explicitly involve particle mechanics. This and the interpretation of \ref{FC} as Newton's equation of motion (despite the comment just before equation \ref{dp/dt=F})     are the principle results of this paper.

\bn
\subsection{Toy Model}
\begin{figure}[H]
\begin{center}
\includegraphics[scale=.4]{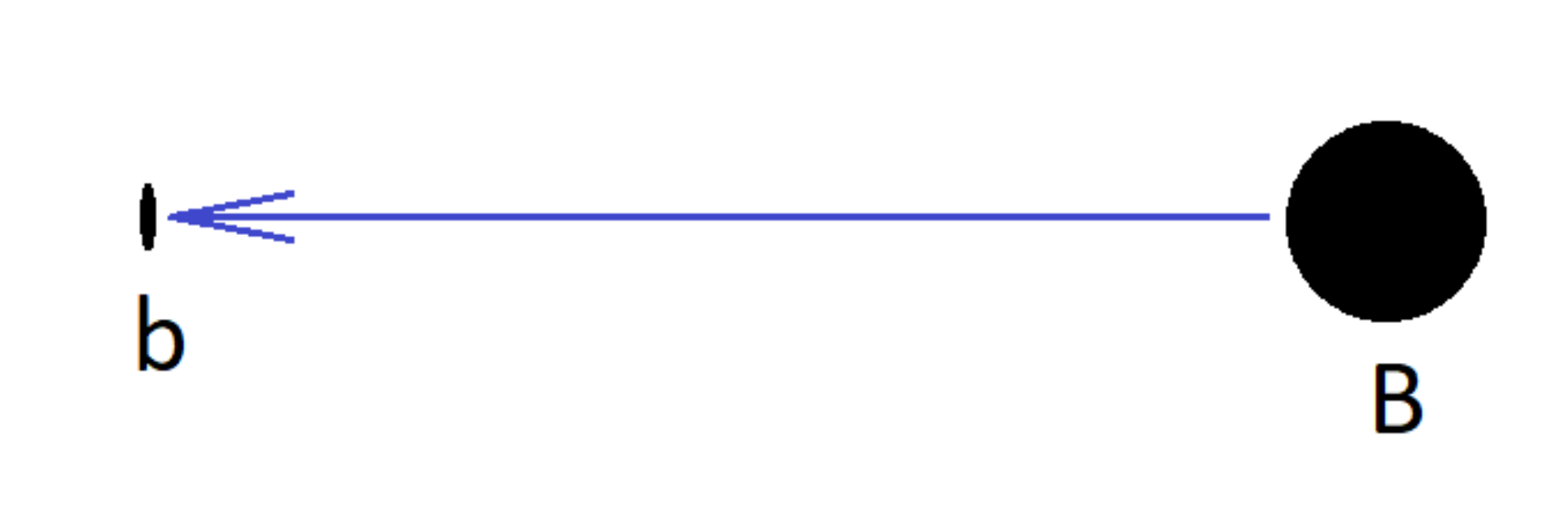}
\caption{Toy model involving a big and little ball. The big ball represents the boundary and little ball represents the particle. The big ball remains nonrelativistic while the little ball quickly become relativistic.}
\label{bigball}
\end{center}
\end{figure}

Equation \ref{FC}  looks temptingly like Newton's equation $F=ma$ for a non-relativistic particle in a uniform gravitational field but for the reason stated above, it does not make sense to identify that particle with the relativistic infalling particle.
To understand what is going on consider a toy   model. Two balls, $\bf{B}$ and $\bf{b}$
are shown in figure \ref{bigball}.
 One---the big-ball $\bf{B}$---is very heavy with mass $M_B$ and the other---little-ball $\bf{b}$---is very light with mass $m_b.$ Initially the two are attached and the combined system is at rest. At $t=0$ the two balls are ejected from one another along the $X$ axis with equal and opposite momentum. We also assume the balls repel each other with a constant force. The result is that  $\bf{b}$ will quickly become relativistic while  $\bf{B}$ remains non-relativistic. Throughout the motion the momenta of the balls are equal and opposite.

It is evident from Newton's third law that both balls satisfy the equations,
\be 
dP/dt =F
\ee
but only  $\bf{B}$ satisfies the non-relativistic Newton equation.
\be 
F = M_{B}\frac{d^2X}{dt^2}.
\ee

The connection between the toy model and the NERN system is  clear:
$\bf{b}$  is the light particle that was dropped from the black hole boundary, and $\bf{B}$ is the boundary itself with mass $M_B.$ 

It is also worth noting that the heavy ball $\bf{B}$ serves as a quantum frame of reference  \cite{Aharonov:1967zza}. As Maldacena has noted, this is similar to the way that
 the condensate of a superfluid or superconductor serves as a frame of reference for a phase variable.
\bn

These considerations, along with equation \ref{FC},
lead  to the conclusion that it is \it  the nonrelativistic  velocity  of the heavy boundary, \rm not the particle,  which is proportional to the rate of change of the complexity of $\psi(t)$, and that it satisfies the Newtonian equation \ref{FC}.

\bn

Since $P$ is conjugate to $\rho,$ and the boundary is non-relativistic, we can write,
\bea
P\eq M_B \frac{d\rho_B}{dt} \cr \cr
\eq \frac{d\CC}{dt}.
\eea
where $\rho_B$ is the location of the boundary. If further follows that,
\be 
\CC =M_B( \rho_B - \rho_0)
\ee
where $\rho_0$ is constant. The obvious choice is for $\rho_0$ to be the horizon location in which case
$\CC$ is proportional to the distance of the boundary from the horizon.   In section \ref{Sec: Formal} where the two-sided case is discussed, the distance defining complexity is naturally taken to be the distance separating the two boundaries.

\subsection{Comparison with CV}

There are a number of ways  of estimating  the boundary mass $M_B.$ One way is to directly analyze the Schwarzian boundary term in the action. I will do something different making direct use of the  
 complexity-volume  (CV) correspondence  \cite{Brown:2015bva}\cite{Carmi:2017jqz};
volume now referring to the length of the throat times its area. For this subsection I will not bother keeping track of numerical factors.

The standard volume-complexity (CV) relation is,
\be 
\CC = \frac{V}{G l_{AdS}}
\label{CV}
\ee
The volume is the area of the throat times the length $\rho,$
\bea
V = A \rho
\eea
where $A$ is the horizon area. Also observe that $A/G $ is proportional to the entropy of the black hole and the  AdS radius is proportional to $ r_+.$
One  finds
\be 
\CC \approx   \lf \frac{S}{r_+}\rg \rho
\ee
or using the SYK/NERN dictionary,
\be 
\CC \approx  \CJ N \rho
\ee
From
 \ref{FC} we may write,
\be 
{
F \approx   \CJ N \frac{d^2 \rho}{dt^2}.
}
\label{F=MA}
\ee
It follows that the mass of the boundary is,
\be 
\boxed{
M_B \approx  
\CJ N .
}
\label{MB}
\ee
This is to be compared with the energy of the infalling particle which is $\CJ$.  The big-ball, little-ball analogy is quite apt. Another point worth noting is that $M_B$ is of the same order as the mass of the NERN black hole.
\be 
M_{BH} = \frac{r_+}{G} \sim \CJ N.
\ee

If we now combine \ref{F=MA} and \ref{MB} with equation \ref{Newt} from the appendix we arrive at Newton's equation,
\be 
\boxed{
 \frac{m_b M_BG}{r^2} = M_B  \frac{d^2 \rho}{dt^2}.
 }
 \label{NewtG}
\ee
for the motion of the boundary\footnote{It should be kept in mind that the $r$ that appears in the inverse square law is not generally the distance of the test particle to the gravitating mass. According to Gauss' law it is the radius of the 2-sphere surrounding the mass at the test point. Only in flat space is it the distance to gravitating mass. }.

The derivation in appendix \ref{App: Particle Equation of Motion} of the left side  of \ref{NewtG} was based on the bulk equation of motion for a  particle in a gravitational field. One may wonder whether it can be derived from the holographic SYK quantum mechanics. The answer is that up to factors of order unity, it can. Using the SYK/NERN dictionary in section \ref{Sec: SYK-NERN} we can write $ \frac{m_b M_BG}{r^2}$ in terms of SYK variables (for $q=4$),

\be
(m_b)(M_B)(G)\lf \frac{1}{r_+^2} \rg = (2\CJ)(M_B) \lf \frac{.12}{N\CJ^2}  \rg \lf  \frac{\CJ^2}{.01} \rg
\label{univ1}
\ee
On the other hand, the right side is just $d^2\CC/dt^2$ which can be evaluated from the Qi-Streicher formula. In the throat region one finds the QS formula gives
\be 
d^2\CC/dt^2 = 4 J^2.
\label{univ2}
\ee
Equating the right side of \ref{univ1} to the right side of \ref{univ2} determines the value of $M_B,$
\be 
M_B\approx .2 N\CJ,
\ee
consistent with  \ref{MB}. 

\bn

There is also information in the Qi-Streicher formula about the relativistic motion of the light particle. For example consider the time that it takes, moving relativistically, for the particle to travel the distance 
$\Delta\rho  = r_+\log{\beta/r_+}$ from the boundary to the Rindler region. From \ref{trajec} one sees that the time is $\beta.$ Once the particle is in the Rindler region the size begins to grow exponentially with time. The Qi-Streicher formula \ref{Qi-Streicher} shows  that this is indeed the case. 

\sc
\section{Formal Considerations}\label{Sec: Formal}

\subsection{Symmetries of $AdS_2$}
The basis for  the derivation of Newton's equations in section \ref{Sec: Newt}  was the relation between momentum and the time derivative of complexity, equation \ref{p=dC/dt}, which itself  was based on
 the momentum-size correspondence  of  \cite{Brown:2018kvn}. The momentum-size correspondence fit
some non-trivial  facts about scrambling by NERN black holes  \cite{Leichenauer:2014nxa},   but it  was never derived from first principles. If we had an alternate route to \ref{p=dC/dt} we could turn the argument around and derive the momentum-size correspondence. Maldacena, Lin, and Zhao  \cite{LMZ} described such a route which I will briefly explain as far  as I understand it\footnote{I am grateful to Henry Lin and Ying Zhao for explaining the argument to me.}. 

\begin{figure}[H]
\begin{center}
\includegraphics[scale=.2]{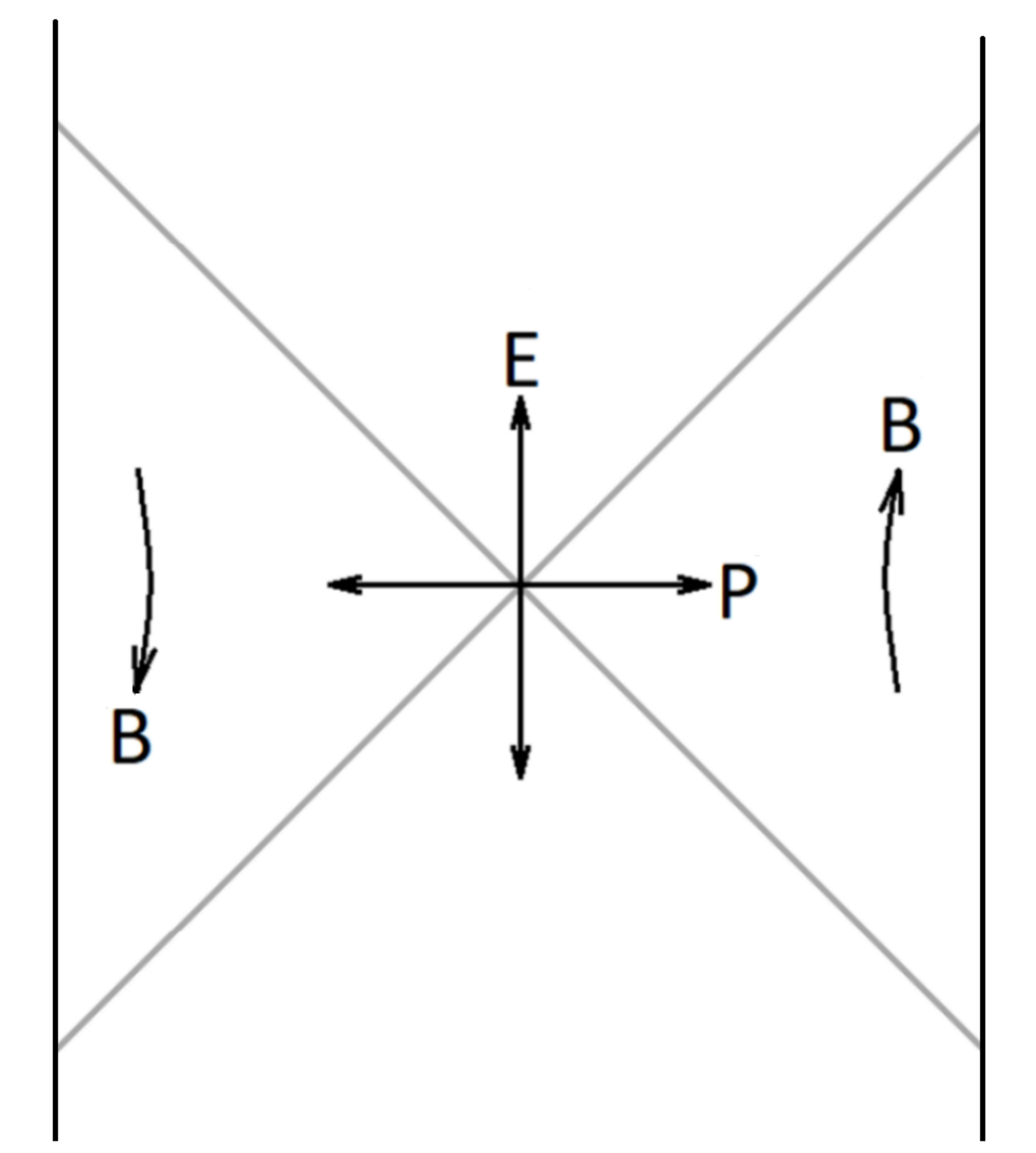}
\caption{Penrose diagram for a two-sided non-dynamical background in the limit of low temperture and infinite throat length. Also shown are the matter generators $E,B, P$ that generate   $SL(2,R)$ motions of the matter fields. The generators have been normalized so that the commutation relations are $[B,E] = iP, \ \ [B,P] = i E, \ \ [P,E] = i\CJ^2 B.$     }
\label{gauge0}
\end{center}
\end{figure}

We begin by considering the approximate symmetries of  matter in the background of a fixed, almost infinite, $AdS_2$ throat. The Penrose diagram for the throat is shown in figure \ref{gauge0}.
The symmetry of infinite $AdS_2$ is the non-compact group $SL(2,R)$. If $\beta$ is finite the symmetry is approximate. Deep in the throat the geometry is indistinguishable from $AdS_2$ but the left and right boundaries break the symmetry. As long as matter is far from the boundaries the symmetry will be respected. 

$SL(2,R)$ has three generators called $E_0, P_0, B_0,$ satisfying the algebra,
\bea
[B_0, E_0] \eq iP_0 \cr \cr
[B_0, P_0] \eq i E_0 \cr  \cr
[P_0, E_0] \eq i B_0
\eea
It is conventient to rescale $P$ and $E$ in order to give them units of energy. Thus define,
\bea 
E \eq \CJ E_0 \cr \cr
P \eq \CJ P_0 \cr \cr
B \eq B_0.
\eea
The commutation relations become,

\be 
\boxed{
[B, E] = iP
}
\ee
\be 
\boxed{
[B, P] = iE
}
\ee
\be 
\boxed{
[P, E] = i\CJ^2 B
}
\ee

Let's consider the generators one by one. The action of $E$ is to shift the Penrose diagram rigidly in the vertical direction. We can introduce a time variable $\tau$ that is constant on  horizontal slices, and which at the center of the diagram registers proper time. $E$ may be represented by the differential operator,
\be 
E = i \frac{\partial}{\partial \tau}.
\ee

The generator $P$ shifts the diagram along spacelike directions. It has fixed points at the asymptotic boundaries on the $t=0$ slice. It may be thought of as the translation generator with respect to the proper coordinate $\rho$  defined in \ref{rho},
\be 
P=-i \frac{\partial}{\partial \rho}
\ee
 Finally $B$ is the boost generator  that has the bifurcate horizon as a fixed point. It is conjugate to the Rindler hyperbolic angle $\omega.$
 \be 
 B = -i \frac{\partial}{\partial \omega}.
 \ee
 
 The Rindler time is related to $t$ by,
 \be 
 \omega = \frac{2\pi t}{\beta}
 \ee
 so that $B$ can be written,
  \be 
 B =-i\frac{\beta}{2\pi} \frac{\partial}{\partial t}
 \ee
 
\begin{figure}[H]
\begin{center}
\includegraphics[scale=.3]{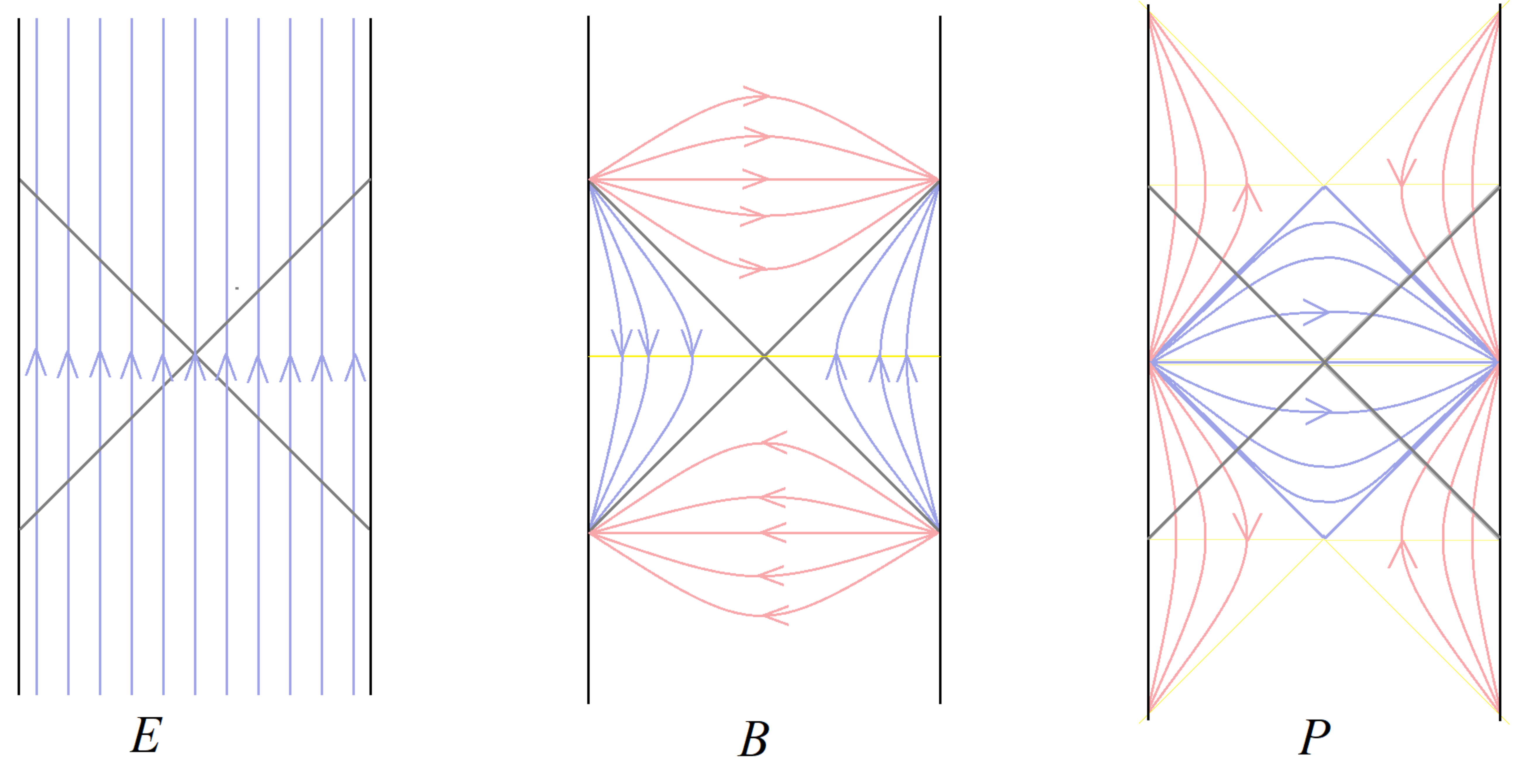}
\caption{Orbits of the three generators.     }
\label{generators}
\end{center}
\end{figure}

 The orbits of the three generators are shown in figure \ref{generators}.

 \bn
The two-sided Penrose diagrams \ref{gauge0} and \ref{generators}  represents two uncoupled but entangled SYK systems with Hamiltonians $H_R$ and $H_L$. The generator $B$ is given in terms of the two Hamiltonians by
\be 
\boxed{
B =\frac{\beta}{2\pi} (H_R - H_L)
}
\ee

\subsection{Left-Right Interaction}
One might think that the 
 global energy $E$ should  be identified with $\beta \CJ [H_L + H_R].$    However, there is no symmetry of $AdS_2$  generated by $(H_L + H_R).$  Without going into details, Maldacena and Qi   \cite{Maldacena:2018lmt} argue that the generator ${E}$ requires the introduction of another term, $H_{int}$  that couples the left and right sides,
\be 
\boxed{
E=\beta \CJ ( H_L + H_R + H_{int}).
}
\ee
Using 
$$ i[B, E] =P $$ and $$B =i\beta \frac{d}{dt}$$ we can write 
\bea 
P&=&i   \beta \CJ   [B, H_{int} ] \cr \cr
\eq \beta^2 \CJ \frac{d H_{int}}{d t}
\label{dHint}
\eea

\bn

In reference \cite{Qi:2018bje} an operator representing size was constructed  in terms of the two-sided degrees of freedom $\psi_{i L}$ and $\psi_{i R}.$ Using our convention of calling size $\CC,$
\be 
\CC = \frac{i}{ \delta_{\beta}} \sum_i \psi_{iL}\psi_{iR} 
\ee
where $\delta_{\beta}$ is a dimensionless normalization factor which normalizes the size of a single fermion to unity.
This same operator appears in the interaction term $H_{int}$ in \cite{Maldacena:2018lmt},

\bea 
H_{int} \eq i \mu   \sum_i \psi_{iL}\psi_{iR}. \cr \cr
\eq \mu \delta_{\beta} \  \CC
\label{Hint=mus}
\eea

Combining \ref{Hint=mus} with \ref{dHint} we find,
\be 
\boxed{
P = \mu \delta_{\beta}\beta^2 \CJ \  \frac{d\CC}{dt}
}
\label{forget}
\ee
Thus, apart from the numerical factor $ \mu \delta_{\beta}\beta^2 \CJ$ one finds that the matter momentum $P$ is indeed proportional to  the time derivative of the size. However, consistency with \ref{p=dC/dt} requires a relation between the parameters
 $ \mu,  \delta_{\beta},   \beta, \rm and \  \it  \CJ, $
\be  
\mu \delta_{\beta}\beta^2 \CJ \approx 1.
\label{1-relation}
\ee

Again, the  meaning of $\approx$ in \ref{1-relation} is: \it equals up to a numerical constant which may depend on $q$. \rm  This is a significant constraint since the parameters $\mu$ and $\delta_{\beta}$ have an intricate mixed dependence   \cite{Maldacena:2018lmt}
on $q$ and the dimensionless quantity $\beta \CJ.$


\subsection{Determining the Prefactor}
It is known that the quantity $\mu$ is not independent of the other three parameters and that there is a relation between them.           Zhao\footnote{Unpublished communication.} has suggested that the coefficient $ \mu \delta_{\beta}\beta^2 \CJ$ can be determined by comparing the calculation of $P(t)$ using the equation of motion in appendix \ref{App: Particle Equation of Motion}, with the Qi-Streicher formula \ref{Qi-Streicher}. From the appendix the force on the infalling particle is constant during passage through the throat and given by  $F \approx \CJ^2.$     It follows that,
\be 
P(t) \approx \CJ^2 t.
\ee
Differentiating the Qi-Streicher fomula also  gives,
\be 
\frac{d\CC}{dt} = 4 \CJ^2 t.
\ee
(In appendix \ref{App: Comparing} a more complete comparison between the particle orbit and the Qi-Streicher formula is carried out for the entire range of $\rho$ from the boundary at $r=2r_+$ to the horizon at $r=r_+.$)

It follows that the coefficient $\mu$ must satisfy,
\be 
\boxed{
\mu \delta_{\beta}\beta^2\CJ \approx 1
}
  \label{mu}
\ee
so that  \ref{p=dC/dt} is satisfied. Equation \ref{mu}   is  non-trivial. On dimensional grounds the $q$ can appear in any combination with the  product $\beta \CJ,$ 
 but \ref{mu} allows only a multiplicative dependence by a function of $q$ alone.

 That the product  in \ref{mu} should only depend on $q$ is non-trivial and is confirmed
in the analysis of \cite{Maldacena:2018lmt} where it appears in a somewhat hidden form in equations $[4.25], \ [4.29],$ and $[4.50].$

The formal considerations of this section did not involve the momentum-size correspondence \ref{p-SIZE} postulated in  \cite{Susskind:2018tei}\cite{Brown:2018kvn} but they  would allow us to work backward from \ref{p=dC/dt} and derive it.

\bn

We are almost where we want to be, but not quite because we have assumed the throat is infinite. If we make the throat finite by allowing $T$ to be small but not zero, the symmetry of the matter system will be broken by the interaction of the matter with the boundary. In a sense that's not surprising since the matter will interact with the dynamical boundary (through the potential barrier) so that the momentum of the matter will not, by itself,  be conserved.

\begin{figure}[H]
\begin{center}
\includegraphics[scale=.3]{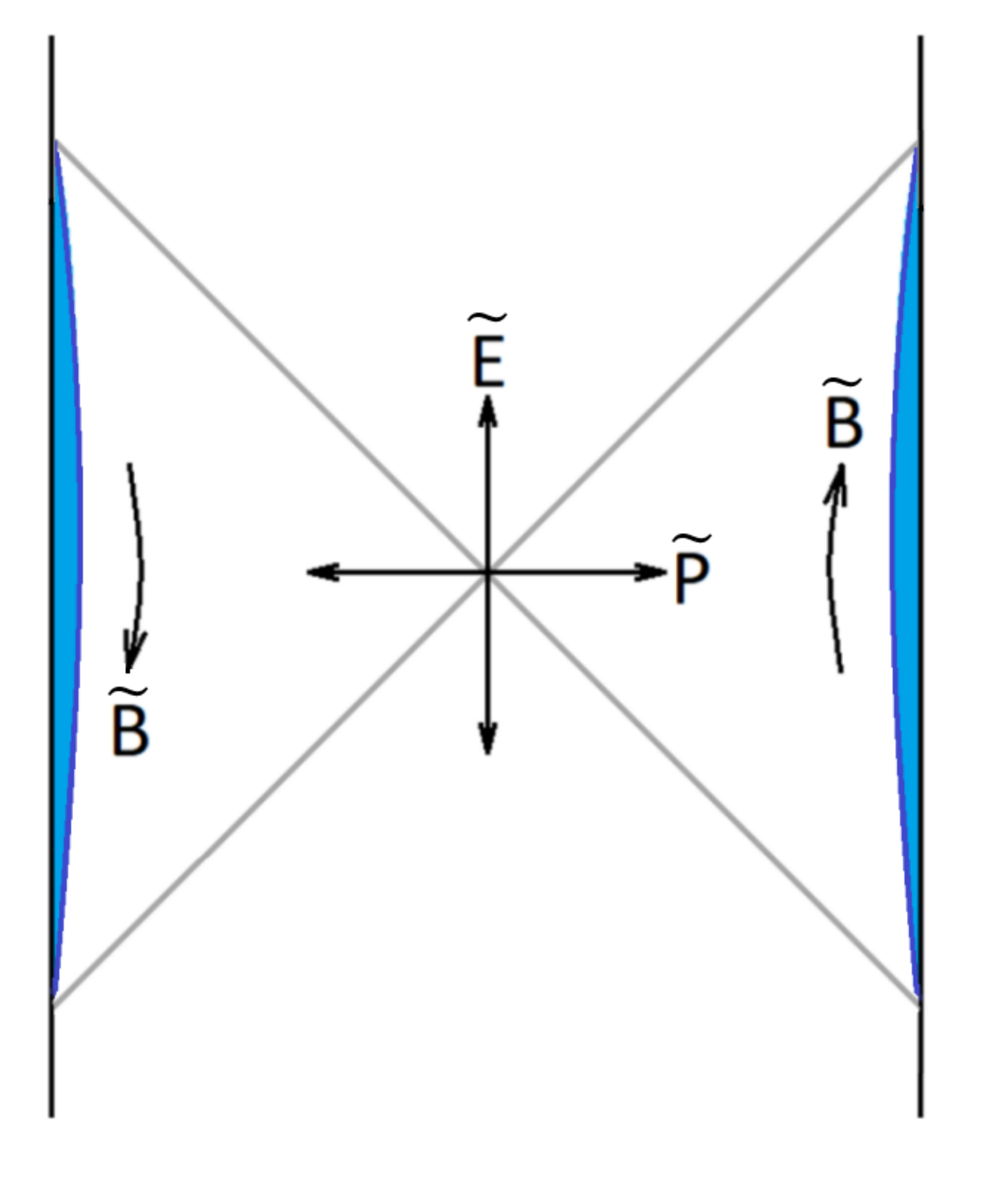}
\caption{Embedding a finite throated geometry in $AdS_2.$ Also shown are the three $SL(2,R)$
gauge generators. The blue regions are part of the embedding space but not part of the actual finite temperature spacetime. The inner boundaries of the blue region are the dynamical boundaries governed by the Schwarzian action.}
\label{gauge1}
\end{center}
\end{figure}

There is a formal way to restore the symmetry as a gauge symmetry  \cite{Maldacena:2016upp}\cite{Yang:2018gdb}\cite{LMZ}. Although the finite throat does not have $SL(2,R)$ symmetry it can be embedded in $AdS_2$ as illustrated in figure \ref{gauge1}.

\bn

The curved  boundary separating the blue regions  from rest of the diagram represents the Schwarzian boundary. The Penrose diagram can be conveniently parameterized by dimensionless coordinates $-\infty < T < \infty$ and $0 < X < \pi$. The embedding is not unique due to the $SL(2,R)$ invariance of $AdS_2$. This invariance allows us to move the geometry in various ways. In other words the representation of the finite throat in $AdS_2$ is redundant; the symmetry is a gauge symmetry. As such its generators should be set to zero. Denoting the gauge generators by tilde-symbols,
\be 
\boxed{
\tilde{E} = \tilde{B} = \tilde{P} = 0
}
\ee
But the tilde generators are no longer the matter charges; they now include the charges of the boundary. In particular the spatial charge $\tilde{P}$ is,
\be 
\tilde{P} = P + P_{boundary}.
\ee
Therefore the gauge condition 
\be 
\tilde{P} = 0
\label{P=0}
\ee
 is the Newtons third law of  action and reaction, which tells us that the boundary recoils when the matter particle is emitted into the throat.
Keeping track of the action=reaction condition seems to be the main point of the gauge symmetry. The un-hatted operators are the physical matter generators and their negatives are the generators that act on the boundary degrees of freedom.

\subsection{Fixing a gauge}

The embedding is not unique due to the $SL(2,R)$ invariance of $AdS_2$. This invariance allows us to move the entire geometry---matter and boundary---in various ways by applying the three gauge generators. 

\begin{figure}[H]
\begin{center}
\includegraphics[scale=.3]{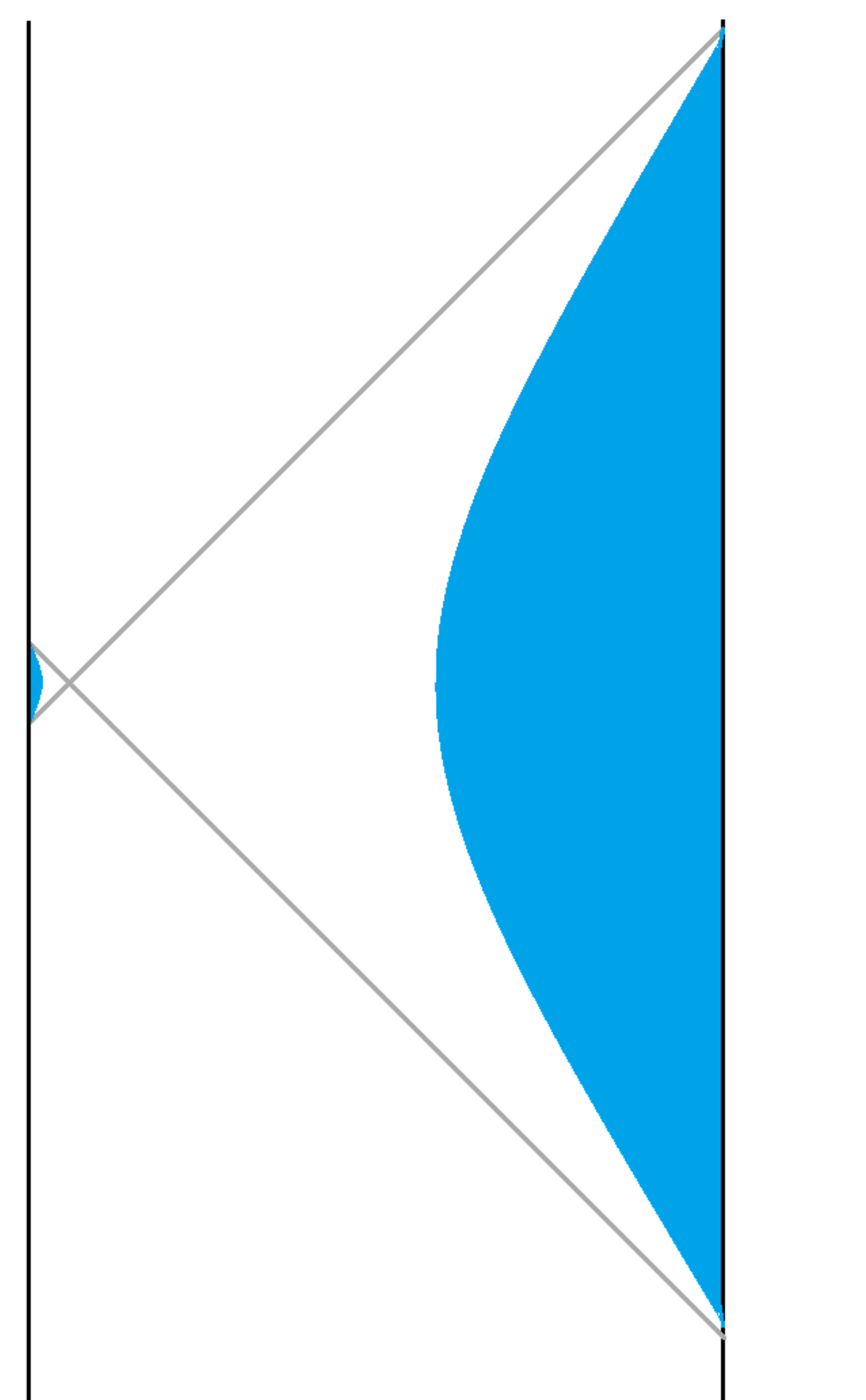}
\caption{Fixing a gauge}
\label{gauge2}
\end{center}
\end{figure}

The action of $\tilde{P}$ moves the bifurcate horizon as well as the excised (blue)  regions. Such a transformation can shift  the NERN geometry from figure [\ref{gauge1}] to [\ref{gauge2}].

 We can use the gauge symmetries them to fix a convenient gauge:
\bi 
\item  The left black hole has a bifucate horizon. Using the $\tilde{E}$ symmetry we can shift it to the $t=0$ slice.
\item  Next we can use  $\tilde{P} $ to shift the position of the right boundary so that it passes through the spatial  midpoint of the diagram on the $t=0$ slice.  More generally we can choose a point $X_0$ in along the $t=0$ surface and have the boundary pass through it. This defines a one parameter family of gauges parameterized by $X_0.$

\begin{figure}[H]
\begin{center}
\includegraphics[scale=.5]{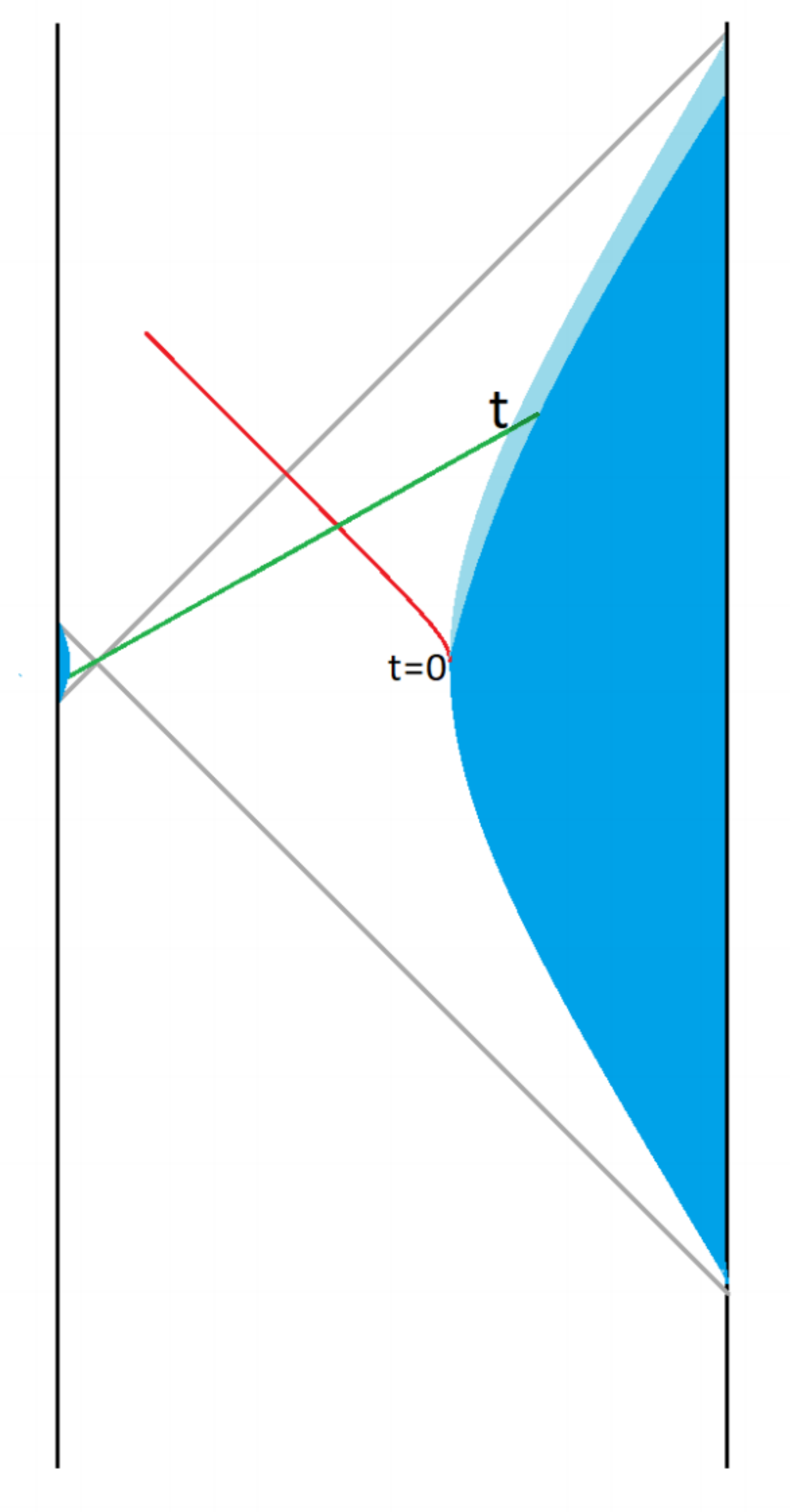}
\caption{The gauge fixed Penrose diagram with the right boundary intersecting the $t=0$ surface at a fixed location half way between boundaries. The red curve is the world line of a particle dropped at $t=0$ from the right boundary. The green surface is boosted from the $t=0$ surface. The boost time $t$ is the time variable that corresponds to the earlier discussion.}
\label{gauge3}
\end{center}
\end{figure}

\item Finally we can fix the boost symmetry by assuming a particle is dropped from the right boundary at $t=0.$
\ei
That completely fixes the gauge. The resulting Penrose diagram is shown in \ref{gauge3}.

Notice that in the limit that that the temperature goes to zero that the bifurcate horizon moves all the way to  the left boundary. The right Rindler patch becomes the Poincare patch, and the boosts become  Poincare time translations. Again there is a one parameter family parameterized by $X_0.$
The boost operator $\tilde{B} $ may now be used to boost the $t=0$ surface forward in time to as illustrated by the green line in figure \ref{gauge3}.

The transformations generated by $\tilde{P}$ are  shifts of the $X_0$ parameter that move the right boundary. 
The momentum of the infalling particle that we called $P(t)$ is the proper momentum on that slice.  

Dropping the particle from the 
 right-side boundary causes the boundary to recoil and move outward. That is indicated by the small separation shown as light blue.  The effect is to change the right-side horizon (not shown)  so that its bifurcate point is no longer on the $t=0$ surface but is slightly below it. The bifurcate point on the left horizon is unchanged.

 The time-slice shown as green is anchored on the boundaries at ``boost time" $t$. The holographic quantum system---two copies of SYK---has a quantum state associated with the time slice and if the particle had not been thrown in, the state would be independent of the time $t$. But the insertion of $\psi_R$ at $t=0$ breaks the boost symmetry and the state evolves with $t$. Since $\psi_R$ is a purely right-side operator it evolves according to,
\bea 
\psi(t)&=&  e^{-i(H_R - H_L)t} \psi  e^{i(H_R - H_L)t} \cr \cr
\eq e^{-iH_R t} \psi  e^{iH_R t}.
\eea  
Under this evolution $\psi_R(t)$ grows in the way I described earlier.

The complexity of the evolving state can be determined from  $CV$ duality. Apart from some constant factors it is just the length of the geodesic connecting the left and right boundaries at time $t$.
If the particle had not been thrown in, the boost symmetry would imply that the length/complexity would be constant, but the small kick causes the length/complexity to grow after the particle is dropped in.

Lin-Maldacena-Zhao argue that the generators can be decomposed into bulk matter, and gravitational (boundary) contributions. The bulk matter contribution to $\tilde{P}$ is the momentum $P$. In the case in which a particle has been dropped into the geometry, $P$ is the particle's momentum.  The gravitational part on the other hand  is the momentum of the heavy non-relativistic boundary, which by the gauge condition is $-P$. (In the case at hand only the right boundary recoils. The momentum of the left boundary stays zero.) The fact that the sum of the particle and boundary momentum must be zero is Newton's third law of action and reaction.

The low energy  $SL(2R)$ symmetry of SYK dictates a particular form for the action governing the motion of the boundary. Known as the Schwarzian action, it is equivalent to the
 Gibbons-Hawking-York extrinsic curvature  that has to be added to the Einstein Maxwell action in the presence of boundaries.  It's rather complicated but in the non-relativistic limit when the boundary moves slowly, the kinetic term in the Schwarzian action must reduce to the action for a non-relativistic particle\footnote{I am grateful to Herny Lin for a helpful discussion of this point.} of mass $M_B=N \CJ$, or in NERN terms, $M_B=S/r_+.$
\be 
I \approx \frac{1}{2}M_B  \  \dot\rho^2.
\ee
This agrees with the analysis in the previous section and provides a formal justification for it.

\bn

In addition there is a coupling between the matter and the boundary which has the form of a repulsive potential energy. As long as the particle is in the throat region the potential is  linear in the distance between the infalling particle and the boundary. As shown in the appendix this leads to a constant  Newtonian force which accelerates both the particle and the boundary in opposite directions, so as to keep the total momentum zero. The result is that  the particle is effectively attracted toward the horizon, and as it falls the complexity grows according to the pattern described in earlier sections and in appendix \ref{App: Particle Equation of Motion}.

\sc
\section{Falling Through Empty $\bf AdS_2 \it $}\label{Sec: empty}

References \cite{Susskind:2018tei} \cite{Brown:2018kvn}, and the present  paper up to this point, deal with the gravitational attraction of a black hole. If the tendency for complexity to increase is  the general holographic mechanism behind gravitation it is important to demonstrate it outside the black hole context.
For example we would like to know when a particle falls toward an ordinary cold mass with little or no entropy, does the holographic  complexity grow? What happens when a comet falls in a long elliptical orbit toward the sun and then goes off into interstellar space. Does the complexity increase and  decrease periodically? 

We could try modeling questions like this in AdS/CFT, but the tools I've used in this paper are special to SYK. Fortunately there is a simple case in which the question can be addressed. Anti de Sitter space has a gravitational field even in the AdS vacuum. The negative vacuum energy of AdS gravitates and attracts matter to the center. One does not need an additional mass. 

The metric of AdS is,
\bea 
ds^2 \eq -f(r)dt^2 +\frac{1}{f(r)} dr^2 + r^2 d\Omega^2  \cr \cr
f(r) \eq \lf 1+\frac{r^2}{l_{AdS}^2} \rg
\eea
Particles dropped from a distance experience an attractive radial gravitational force which behaves similarly to a harmonic oscillator force. A particle will move in a periodic orbit oscillating about the origin. There is no black hole, no horizon, no entropy. 

\begin{figure}[H]
\begin{center}
\includegraphics[scale=.2]{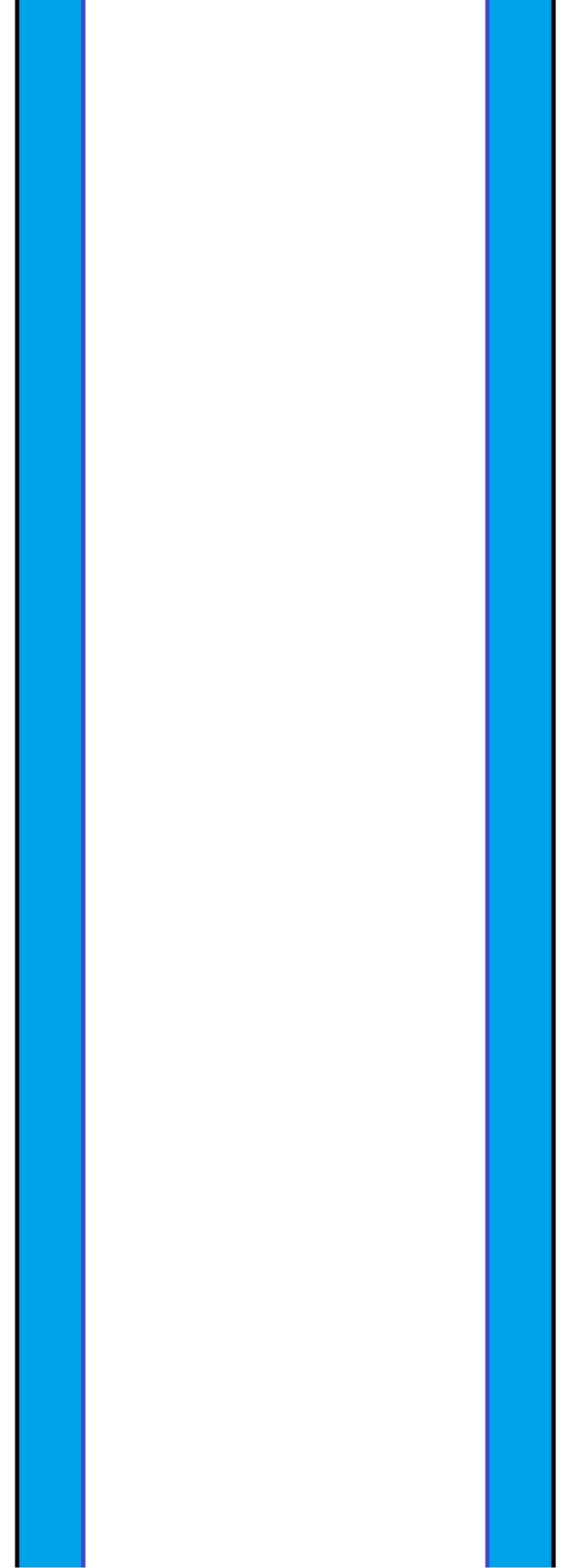}
\caption{Traversable wormhole with two boundaries.}
\label{ads}
\end{center}
\end{figure}

Two dimensional AdS is not an exception, but engineering empty $AdS_2$ is subtle in the SYK system.
Maldacena and Qi \cite{Maldacena:2018lmt} arrange it by perturbing a two-sided black hole with a Left-Right interaction. The resulting space is called a traversable wormhole;  in fact it is a cutoff version of $AdS_2$. The geometry does not extend out to $r=\infty,$ but instead is cut off at some large radial distance by a Schwarzian boundary, or to be precise, two Schwarzian boundaries\footnote{AdS two is unique in having two disconnected boundaries.}: one for the left side and one for the right side, as in figure \ref{ads}.  The geometry is $AdS_2$ except that the blue regions near the boundary have been excised. 
\bn

\begin{figure}[H]
\begin{center}
\includegraphics[scale=.3]{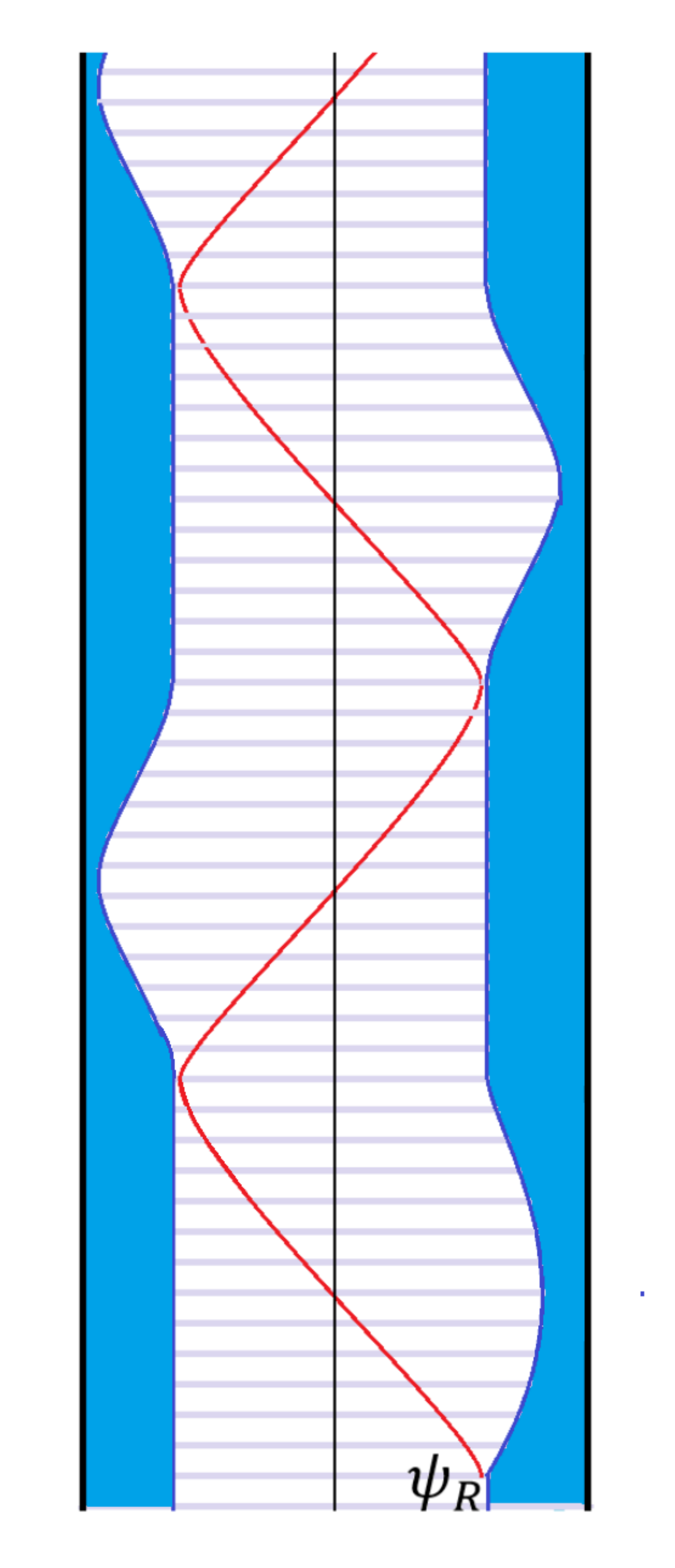}
\caption{A particle has been added to the right side of the  traversable wormhole by acting with $\psi$.
The subsequent motion is oscillatory with periodic variation in the distance between the boundaries, thus indicating periodic variation of complexity. 
The figure has been foliated with constant time slices to help guide the eye.  The oscillations of the boundary are very small and have been greatly  exaggerated.}
\label{viggle}
\end{center}
\end{figure}

 In figure \ref{viggle}, by applying a right-side fermion operator  a particle can be dropped in from the right boundary. The initial state has the form 
 $$\psi_R |0\ra$$
 and subsequently evolves to $$\psi_R (t) |0\ra.$$

  In this case there is no black hole and the particle endlessly oscillates back and forth between the two boundaries.
The force on the particle is gravitational. From the bulk GR viewpoint  it is produced by the vacuum energy in the region between the boundaries. 
The state without the particle (figure \ref{ads}) is the ground state of the Hamiltonian  and the complexity---in this case represented by the distance between the two boundaries---is constant in time. 

When the particle is injected at $t=0$ by applying the right-side fermion operator $\psi_R$ the additional complexity of the state is initially  very small.  As the particle accelerates toward the center of  AdS  its momentum increases. The right boundary recoils  so that the distance between the boundaries increases. According to  CV duality, the complexity also increases.

Because the boundary is very heavy it moves non-relativistically which means its momentum and velocity are proportional to one another, and once again,
\be 
P \approx \frac{d\CC}{dt}
\ee
for both the boundary and for  the particle.

The radial momentum reaches a maximum when the particle reaches the center of the diagram. It then switches sign. At the same time the complexity starts to decrease\footnote{This conclusion is based on the ability of the gravitational dressing to switch from the right to the left side. Such switching would be impossible without  left-right coupling, but there is no obstruction to it when the Maldacena-Qi interaction is included.}.
By the time the particle reaches the left boundary the complexity has decreased to its original value. The state at that point is $$\psi_L|0\ra.$$

 The particle then gravitates back to the center and subsequently returns to the right boundary.
The oscillating behavior of complexity may seem odd, but in fact it is generic for integrable systems. It is also characteristic of holographic systems below the black hole threshold \cite{Anous:2019yku}.

To reiterate, the connection between gravitational attraction and complexity is not dependent on the presence of a black hole, or on the presence of  a system with a large entropy. However without a black hole the  system is integrable and    the complexity oscillates. It should be pointed out that the complexity never get's very large during the oscillating behavior. At the maximum when the particle is at the center of the geometry the complexity is $\sim \beta^2 J^2$ which is much less than $N,$ i.e., the complexity at scrambling.

It would be interesting to confirm this behavior in the SYK theory using the methods of Qi and Streicher.


\sc
\section{Concluding Remarks}

In this article I have assembled further  evidence that the holographic avatar of gravitational attraction is
 the growth of operator-size during the run-up to the scrambling time. During this period, size and complexity are indistinguishable, and one can say that gravitational attraction is an example of the tendency for complexity to increase.
 The presence of a massive object creates a kind of complexity-force, driving the system toward greater complexity in the same way that an ordinary force accelerates a particle toward lower potential energy.
This conclusion was based on three things: the CV correspondence  between complexity and volume; 
 a  duality between momentum and the time-derivative  of complexity,
$$
P \approx \frac{d\CC}{dt};
$$
and the Qi-Streicher calculation of the time dependence of size at low temperature.

To test the duality, on the left side we used the standard relativistic classical theory of particle motion (in a gravitational field) to compute $P(t).$  On the other side the Qi-Streicher calculation of $\CC(t)$  (a pure quantum calculation that makes no reference to particle motion) allows us to compute $\frac{d\CC}{dt}$. The two sides agree.

\bn

One can object to such a  connection (between momentum and complexity) on the grounds that  it relates two fundamentally different kinds of quantities. Momentum is a linear quantum observable. Complexity is a nonlinear property of states; linear superpositions of states with the same complexity may have very different complexity. Thus equating momentum and the time-derivative of complexity is inappropriately mixing concepts. 

Similar things have been seen before. The Bekenstein formula and more recently, the Ryu-Takyanagi formula, equate area---a quantum observable---to entropy. This also seems inadmissible for similar reasons. A number of authors have written about this tension  (see for example \cite{Papadodimas:2015jra}\cite{Harlow:2016vwg} and references therein)
and the resolution seems to be that quantities like entropy may behave like  observables over a relatively small subspace of states---a so called code subspace.  Thus, for states near the ground state of AdS, area and entanglement entropy can coincide, but the relation does not hold for most states.  

The same things should be true for complexity: in the small subspace of states  encountered while a particle is falling toward the horizon of a black hole complexity and its derivative can behave like an observable, but beyond the scrambling  time  or when superpositions of classical states are considered the relation between complexity and  observables must break down\footnote{I am grateful to Daniel Harlow for discussions about this point.}.

\bn

  On another point, E. Verlinde  has also  emphasized  the need for a holographic explanation of gravitational attraction and has
proposed an entropic mechanism \cite{Verlinde:2010hp}.  He argues that lowering an object toward a  horizon  increases the thermodynamic  entropy    an entropic force. What I find unclear is how  an entropic mechanism  can explain the gravitational pull-to-the-center in cold empty AdS, or to a conventional zero temperature massive body  in its (non-degenerate)  ground state.  How can an entropic theory be compatible with the periodic oscillations of the distance between the sun and a comet in an elongated orbit?

In  contrast to coarse-grained thermal entropy, complexity and operator size can oscillate, especially for non-chaotic or weakly chaotic systems. By the complexity-volume correspondence, the oscillating complexity may manifest itself as periodic motion.
The motion of a particle in empty $AdS_2,$ discussed in section \ref{Sec: empty} is an example.

Returning to  the case of a black hole, entropy approaches its maximum value well before the scrambling time, but as shown in   \cite{Susskind:2018tei}   and  \cite{Brown:2018kvn}, under the influence of gravity,  the infalling momentum increases exponentially until the scrambling  time has been reached. Again it is not obvious how an entropic theory would deal with this.

\bn

It is quite possible that these remarks represent my own misunderstanding of Verlinde’s theory.

  \section*{Acknowledgements}
  
 I thank Henry Lin and Ying Zhao  for very helpful discussions about both  the heuristic and formal arguments in this paper. The paper would not have been written without the many discussions I had with Alex Streicher, in which he explained his results on the growth of size in SYK, and related issues.
 
  This research was supported by  NSF Award Number 1316699.

\section*{Appendix}

\appendix

\section{Particle Equation of Motion}\label{App: Particle Equation of Motion}

Let's consider the radial motion of a particle of mass $m$ moving in a metric,
\be 
ds^2 = -f(r) dt^2 + d\rho^2
\label{metric}
\ee

The standard Lagrangian is,
\be  
\CL = -m\sqrt{f(r) -\dot{\rho}^2}
\ee
and the momentum conjugate to $\rho$ is given by,
\bea  
P \eq \frac{\partial \CL}{\partial \dot{\rho}} \cr \cr
\eq 
\frac{\dr}{\sqrt{f(r)- \dr^2}}
\eea
The Hamiltonian satisfies,
\be 
H = P \dr - \CL = \frac{mf}{\sqrt{f(r)- \dr^2}}
\label{H}
\ee

The radial gravitational force on the particle is,  
\bea 
  F \eq  \frac{\partial \CL}{\partial \rho} \cr \cr
  \eq \frac{m}{2}  \frac{\partial_{\rho}f}{\sqrt{f-\dr^2}} \cr \cr
  &=& 
   \frac{m}{2}  \frac{\partial_{r}f}{\sqrt{f-\dr^2}} \frac{dr}{d\rho} \cr \cr
   \eq     \frac{m}{2}  \frac{\partial_{r}f}{\sqrt{f-\dr^2}}\sqrt{f}
\eea
  
 Now  using \ref{H}
  \be 
  F= \frac{\partial_r f}{2\sqrt{f}}  H
  \label{force}
  \ee
  
Throughout the passage through the long throat (but not into the Rindler region)  the metric may be approximated by the extremal metric,
\be  
f(r) =\lf 1 -\frac{r_+}{r}      \rg^2
\ee
giving,
   \be 
  F=\frac{r_+H}{r^2}
  \ee
For a particle of energy $\sim 1/r_+ (= \CJ)$ the product $r_+H$ equals $1$ and,
\be 
F = 1/r^2.
\ee

  For most of the passage the radial coordinate $r$ has negligible variation, and we may write
    \be 
  F=\frac{1}{r_+^2} \sim  \CJ^2
  \label{F=JJ}
  \ee

Using Lagrange's equations of motion,
  \be 
  \dot{P} = F
  \ee
we see that while in the throat, the particle moves under the influence of  a constant force. The momentum increases linearly with time,
\be 
\boxed{
P = Ft \sim \CJ^2 t.
}
\label{P=JJt}
\ee

Equation \ref{F=JJ} has a simple significance. From the SYK/NERN dictionary in section \ref{Sec: SYK-NERN} one sees,
\bea
m_b  &\leftrightarrow& \CJ \cr \cr
M_B  &\leftrightarrow& \CJ N \cr \cr
G  &\leftrightarrow& \frac{1}{\CJ^2N} \cr \cr
\frac{1}{r_+^2 } &\leftrightarrow&  \CJ^2.
\eea
Equation \ref{F=JJ} can be rewritten,
$$
F= \frac{m_b M_BG}{r_+^2}.
$$
Recalling that $r$ is very close to $r_+$ throughout the throat, we can express this in the familiar Newtonian form,
\be 
\boxed{
F= \frac{m_b M_BG}{r^2}
}
\label{Newt}
\ee
where $m_b \sim \CJ$ is the energy of the single fermion created by $\psi,$  and  $M_B \sim N\CJ$ is the mass of both the boundary (and  the NERN black hole).

\section{Relativistic Orbit}\label{App: Relativistic Orbit}
Let us consider the trajectory of a massless particle from the start at $r=2r_+$ all the way to the horizon at $r=r_+.$ The light-like trajectory is given by,
\bea 
-dt &=& \frac{1}{f(r)}dr \cr \cr
\eq \frac{r^2 dr}{(r-r_+)(r-r_-)}.
\eea
with boundary condition that $r(0) = 2r_+.$  One finds that for $r< 2r_+$ the solution quickly tends to,
\be 
r =   \frac{ e^{\frac{4\pi t}{\beta}} r_+ -r_-   }{ e^{\frac{4\pi t}{\beta}} -1}        
\ee

\be 
r =  \frac{ x^2 r_+ -r_-}{x^2 -1}  
\label{r(x)}
\ee
where $x$ is defined to be,
\be 
x= e^{\frac{2\pi t}{\beta}}.
\label{x}
\ee
Two useful relations are,
\bea 
 r-r_+  \eq \frac{1}{x^2-1} \delta r  \cr \cr
 r-r_- \eq  \frac{  x^2}{x^2-1} \delta r
 \label{useful}
\eea
where $\delta r = ( r_+ - r_-).$

\section{Comparing Trajectory with Qi-Streicher}\label{App: Comparing}
One can explicitly check the equation $P=d\CC/dt$ using the equations of motion of the particle for the left side and the Qi-Streicher fomula for the right side. A slightly more efficient procedure is to time-differentiate both sides,
$$\dot{P} = d^2\CC/dt^2,$$
and then use the force \ref{force} for the left side, and the Qi-Streicher formula for the right side. Thus we wish to check the following:
\be 
    \frac{\CJ}{2}              \frac{\partial_r f}{\sqrt{f}}  \ 
    {\stackrel{?}{=}} \ 
     \frac{d^2\CC}{dt^2}
     \label{eqmotforC}
\ee
with $$f=\frac{(1-r_+)(1-r_-)}{r^2}.$$
Finally we may use  \ref{r(x)}, \ref{x}, and \ref{useful} to re-express the the left side  as a function of $t.$
For the right side we use the Qi-Streicher formula and differentiate it twice. 

Explicit evaluation is straightforward and (up to the usual constants) gives the same answer for both sides, namely,
\be 
    \frac{\CJ}{2}              \frac{\partial_r f}{\sqrt{f}} \approx \frac{d^2\CC}{dt^2} = 2\CJ^2 \frac{(x^2+1)}{x}
\ee
The relation extends over the entire range of $r$ from $r\sim 2r_+$ to the horizon at $r=r_+.$  In the throat where $x$ is close to $1,$
\be 
\frac{d^2\CC}{dt^2} = 4\CJ^2 
\ee
but in the Region where $x$ becomes large,
\be 
\frac{d^2\CC}{dt^2} = 2x\CJ^2 .
\ee
This relative factor of $2$ between the throat and the Rindler region is the same factor that occurred in equation \ref{tb-dC/dt=s}.

Having checked \ref{eqmotforC} we may integrate it and confirm  the precise agreement between the momentum of the falling particle and the time derivative of $\CC.$

\end{document}